\documentclass[prl,twocolumn,floatfix,superscriptaddress]{revtex4}
\usepackage{graphicx,amsfonts,amssymb,amsmath, hyperref,multirow,array}
\usepackage[utf8]{inputenc}

\newif\ifhyper
\hypertrue
\ifhyper
\hypersetup{
  citecolor = {green},
  colorlinks = {true}, 
  urlcolor = {blue} 
}
\fi

\newlength{\ldag}
\begin{document}

\title{The Liquid-Gas Transition in Granular Matter~: a Question of Effective Friction~?}

\author{O. Coquand} 
\email{oliver.coquand@univ-perp.fr}
\affiliation{Laboratoire de Modélisation Pluridisciplinaire et Simulations, Université de
Perpignan Via Domitia, 52 avenue Paul Alduy, F-66860 Perpignan, France}





\begin{abstract}
	This work presents a comparative study of the best models available to describe granular fluids in order
	to investigate the extent to which it makes sense to speak about a liquid-gas transition in a system
	of particles that present no attractive interactions.
	It is shown that the gas and the liquid correspond to regimes with clearly distinct rheological responses.
	A microscopic interpretation of what happens at the transition in terms of the time scales relevant to the various
	physical processes is also presented and put the test against numerical data.
	Our work calls for more experiments to test our predictions on real systems.
\end{abstract}

\maketitle

\section{Introduction}

The term "\textit{granular matter}" describes all systems which elementary components have a size typically bigger than $100$ $\mu$m;
	They are therefore not showing Brownian motion, and dissipate energy whenever they collide.

	It is usually accepted that granular systems can be classified into three states~: solid, liquid and gas \cite{Andreotti13}, as for
	usual matter.
	However, the simplest, and most widely used models of granular systems consist of a collection of inelastic hard spheres, which
	do not present any kind of attractive interaction.
	Thus, it is natural to wonder about the nature of the transition between a \textit{granular gas} and \textit{granular liquid} states,
	that cannot be directly compared to its equivalent in Brownian systems.

	First, it is important to understand that, because of the dissipative nature of the interactions between granular particles, a fluid state
	(be it gaseous or liquid) can only be maintained if energy is injected into the system.
	This sheds light on another fundamental difference between Brownian and granular systems~: the latter can only be at best in an out of equilibrium
	steady state.
	That is the reason why the thermodynamic term "\textit{phase}" is avoided here to refer to the different classes of granular fluids.

	On top of the above difficulties, granular fluids lack integrative models, that enable to describe both the liquid, and the gas state
	within a unified framework.
	As a matter of fact, the most successful models of granular gases are based on modified versions of the Boltzmann equation
	\cite{Lun84, Lun85, Lun87, Dufty97, Brey98, Montanero99, Andreotti13, Santos98, Gonzalez19, Takada20} that take into account properties of the
	inelasticity of collisions, such as the loss of symmetry under time reversal.
	Such models lead to heavy equations that generally require further approximations to be solved.
	We do not describe all the different models here, a review of the principal ones will be given in the following.
	The main point we want to stress here is that, although such models proved to be robust at quite high packing fractions, the description of the dynamics
	in terms of a collisional operator, inherited from the Boltzmann equation, prevents such approaches from being able to deal with the
	highly collective dynamics of the granular liquid state.

	On the other hand, a model of the dynamics of the granular liquid, based on tools used to describe the physics of dense liquids and colloidal suspensions
	(the Mode-Coupling approximation (MCT) \cite{Goetze08}, and the Integration Through Transients (ITT) approach \cite{Fuchs02,Fuchs09}), called the
	Granular Integration Through Transients approach (GITT), has yielded
	results in good agreement with experiments and simulations in the liquid state \cite{Kranz13,Kranz18,Kranz20,Coquand20f,Coquand20g,Coquand21}.
	As we are going to discuss, such a framework is based on a continuous medium modelisation of the granular packing, which should break down at some
	point in the dilute limit of the gaseous state.

	The purpose of this work is to compare the two classes of models, in particular in the region where the transition between the liquid and the gas
	state should occur, to determine a criterion which discriminates the two kinds of fluids.
	We focus on macroscopic observables related to the rheology of the granular medium, that is how it deforms under shear.
	For sake of simplicity, we restrict ourselves to the study of simple shear flows.
	We also restrict our study to the Bagnold problem, where shear heating is the only source of energy injection and dissipation occurs only
	through particle collisions.

	Macroscopic observables are interesting because they can show universality, that is independence to the details of our models at the microscopic scales.
	Quantities of particular importance will be the packing fraction of the granular state $\varphi$, its pressure $P$, the applied shear stress $\tau$,
	and the so-called \textit{effective friction coefficient} $\mu = \tau/P$ \cite{GDR04}.
	Indeed, one particularity of the granular liquid state is the existence of a very robust link between $\mu$ and the dimensionless shear
	rate $\mathcal{I}=\dot\gamma t_{ff}$ (where $\dot\gamma$ is the shear rate, and $t_{ff}$ the free fall time scale of a granular particle),
	called the inertial number.
	This law is called the $\mu(\mathcal{I})$ law, and goes as follows~:
	\begin{equation}
	\label{eqMuI}
		\mu(\mathcal{I}) = \mu_1 + \frac{\mu_2 - \mu_1}{1 + \mathcal{I}_0/\mathcal{I}}\,,
	\end{equation}
	where $\mu_1$, $\mu_2$ and $\mathcal{I}_0$ are constants.
	This law is robust enough for the ability of a model to retrieve it to be a necessary condition for it to be an acceptable model of the granular liquid state.
	The GITT model has passed this test \cite{Coquand20f,Coquand20g}.
	It has even received experimental support in a recent study \cite{Angelo23}.

	One particular property of Eq.~(\ref{eqMuI}) is that it yields a monotonous behaviour of $\mu$ as a function of $\mathcal{I}$, and as a function of $\varphi$.
	We are going to show that this monotonicity breaks down at some point when $\varphi$ is decreased, which marks the end of the $\mu(\mathcal{I})$ regime,
	and therefore constitutes a reasonable criterion for the identification of the liquid-gas transition in granular fluids.
	Furthermore, the compared analysis of the major models of granular fluids will provide more arguments to defend this assertion.

	Finally, we will also show that the gas state can be split into two different gaseous regimes with different rheologies~: a dilute gas state, where binary
	collisions dominate, and a moderately dense gas state, where collisions beyond the Stoßzahl ansatz become important.

	The paper is organised as follows~: In a first section, we discuss the major models of the granular gas state. We compare them, and analyse their similarities
	and differences. Then, we discuss the GITT model, and in particular, its low density limit, which has not been exposed in the literature so far.
	Finally, we proceed to the detailed comparison of the different models, and deduce a reasonable scenario for the evolution of $\mu$ with $\varphi$,
	and its consequences in terms of microscopic physics.

\section{Granular Gas Models}

	\subsection{General properties}

	The most successful models to describe granular gases are based on a modified version of the Boltzmann equation, that can
	be written in the following form \cite{Montanero99}~:
	\begin{equation}
	\label{eqBo}
	\left[\frac{\partial}{\partial t} + \mathbf{v}\cdot\mathbf{\nabla}\right]f^{(1)}(\mathbf{r}, \mathbf{v}, t) = \mathcal{J}_E\big[\mathbf{r}, \mathbf{v}\big| f\big]\,,
	\end{equation}
	where $f^{(1)}$ is the single particle distribution function, and $\mathcal{J}_E$ the modified collision operator.

	The next step is to replace the functional collision operator by some physically inspired ansatz.
	Most of those models belong to the class of Revised Enskog Theories (RET) \cite{Beijeren79}, where the collision operator takes the following form
	\cite{Montanero99}~:
	\begin{equation}
	\label{eqJE}
		\begin{split}
			\mathcal{J}_E\big[\mathbf{r},\mathbf{v}\big|f\big] = &\sigma^2\int d\mathbf{v}_2\int d\Omega\\
									     &\Theta(\mathbf{g}\cdot\hat{\mathbf{\sigma}})
			(\mathbf{g}\cdot\hat{\mathbf{\sigma}})\big[\varepsilon^{-2}f^{(2)}(\mathbf{r}, \mathbf{r}-\mathbf{\sigma},\mathbf{v}',\mathbf{v}_2',t)\\
									     &-f^{(2)}(\mathbf{r}, \mathbf{r}+\mathbf{\sigma},\mathbf{v},\mathbf{v}_2,t)\big]\,,
		\end{split}
	\end{equation}
	where $\sigma$ is the particle's diameter, $\hat{\mathbf{\sigma}}$ is the unitary vector pointing in the direction linking two particles, $\Omega$ is the solid
	angle used to describe $\hat{\sigma}$, $\Theta$ is the Heavyside function, $\mathbf{g} = \mathbf{v} - \mathbf{v}_2$, $\varepsilon$ is the restitution coefficient,
	$\mathbf{v}'$ and $\mathbf{v}_2'$ are the velocities after the collision~:
	\begin{equation}
		\begin{split}
			& \mathbf{v}' = \mathbf{v} - \frac{1 + \varepsilon}{2\,\varepsilon}(\mathbf{g}\cdot\hat{\sigma})\hat{\sigma}\\
			& \mathbf{v}_2' = \mathbf{v}_2 + \frac{1 + \varepsilon}{2\,\varepsilon}(\mathbf{g}\cdot\hat{\sigma})\hat{\sigma}\,,
		\end{split}
	\end{equation}
	and $f^{(2)}(\mathbf{r}_1,\mathbf{r}_2,\mathbf{v}_1,\mathbf{v}_2,t)$ is the two particles distribution function.
	In the RET model, the latter is further approximated by the following expression~:
	\begin{equation}
		\begin{split}
			&f^{(2)}(\mathbf{r}_1,\mathbf{r}_2,\mathbf{v}_1,\mathbf{v}_2,t) \\
			&\simeq \chi\big[\mathbf{r}_1,\mathbf{r}_2\big|\varphi(t)\big]
			f^{(1)}(\mathbf{r}_1, \mathbf{v}_1,t) f^{(1)}(\mathbf{r}_2,\mathbf{v}_2, t) \,,
		\end{split}
	\end{equation}
	where $\chi$ is the value of the pair correlation function (in practice, for hard spheres, it is only evaluated at contact).

	The full derivation of the granular gas rheology from the RET model is too long to be reproduced here ; Therefore, we chose to describe the main
	steps of the derivation, referring the reader to the appropriate papers for details.

	Lastly, a major feature of the Enskog models is that they are expansions in small gradients around the equilibrium distribution
	(which takes a Maxwell-Boltzmann form in this case).
	Indeed, almost nothing is known about the exact distribution functions of particles in a sheared system, let alone a dissipative system.
	The Enskog approach proceeds by assuming that, although the distributions are different than those at equilibrium, the first correction to the latter
	correspond to the arising of local inhomogeneities within the medium.
	Therefore, the main coarse grained quantities, such as the density or temperature fields, should develop small local variations.
	The Enskog series is then organized in terms of the order of such local gradients (although in practice, development beyond the next to leading order
	are rarely used).

	The above discussion applies to all the Enskog models of granular gases we are going to discuss.
	We are now going to describe these models more specifically.
	Be aware that, in order to keep consistent notations in this article, some symbols and prefactors may vary compared to the original papers.

	\subsection{The Dilute Elastic Model}

		The first model we are going to discuss is that of Lun et al. \cite{Lun84,Andreotti13}.
		It is based on an expansion of the Enskog equations in terms of the Savage $R$-parameter, defined in \cite{Savage81}
		as the (dimensionless) ratio between the rate of variation of the velocity imposed by the shear and the root mean squared velocity~:
		\begin{equation}
			R = \sigma\,\frac{d u/dy}{\left<v^2\right>^{1/2}}\,,
		\end{equation}
		where $y$ is the depth of the granular packing, $u$ its mean velocity profile, and $v$ the microscopic velocity.
		A small value of $R$ corresponds to flows for which the main origin of velocity fluctuations is not the applied shear \cite{Savage81}.

		Let us try to get a more precise idea of the consequences of assuming $R\ll1$.
		In a unit system where the mass of the particles is $m=1$, $R$ can be expressed as follows~:
		\begin{equation}
			R \propto \sigma\,\frac{\dot\gamma}{\sqrt{T}}\,.
		\end{equation}
		Moreover, the energy conservation equation writes~:
		\begin{equation}
			\tau \dot\gamma = n\,\Gamma_d\,\omega_c\,T\,,
		\end{equation}
		where the right hand side contains the shear heating power, and the left hand side is the dissipated power, expressed as a function
		of the density $n = 6\varphi/\pi \sigma^3$, the dissipation rate $\Gamma_d = (1-\varepsilon^2)/3$, the collision frequency $\omega_c=
		24\varphi\chi/\sigma \sqrt{T/\pi}$ (we recall that $\chi$ is the value of the pair correlation function at contact, for which we do not need
		to give a precise value for now), and the granular temperature $T$.

		Defining the dimensionless shear stress $\overline{\tau}$ as~:
		\begin{equation}
			\tau = \overline{\tau}\,\sigma^2\dot{\gamma}^2\,,
		\end{equation}
		we get~:
		\begin{equation}
			R^3 \propto \frac{n\,\Gamma_d\,(\sigma \omega_c/\sqrt{T})}{\overline{\tau}}\underset{\varepsilon\rightarrow1,\varphi\rightarrow0}{\propto}
			\varphi^4(1-\varepsilon)^{3/2}\,,
		\end{equation}
		where for the last step, we anticipated the result for $\overline{\tau}$ (see below for the details).
		The main point we want to emphasize here is that the $R\ll1$ expansion corresponds both to the dilute and the elastic limit.
		It should thus apply to dilute gases of nearly non-dissipative particles.

		We do not reproduce here the details of the computation.
		The results are as follows \cite{Lun84,Andreotti13}~:
		\begin{subequations}
			\begin{equation}
				r(\varepsilon) = \frac{1+\varepsilon}{2}
			\end{equation}
			\begin{equation}
				F_1(\varphi,\varepsilon) = \varphi + 4\,r(\varepsilon)\varphi^2\chi
			\end{equation}
			\begin{equation}
				\begin{split}
					F_2(\varphi, \varepsilon) = \frac{5\sqrt{\pi}}{96}&\left( \frac{1}{r(\varepsilon)(2-r(\varepsilon))\chi}
					+\frac{8}{5}\frac{3r(\varepsilon)-1}{2-r(\varepsilon)}\varphi\right. \\
												& \left.+\frac{64}{25}r(\varepsilon)\chi
						\left[\frac{3r(\varepsilon)-2}{2-r(\varepsilon)} - \frac{12}{\pi}\right]\varphi^2\right)
				\end{split}
			\end{equation}
			\begin{equation}
				F_3(\varphi,\varepsilon) = \frac{8}{\sqrt{3\pi}}r(\varepsilon)\chi\,\varphi^2
			\end{equation}
			\begin{equation}
				F_5(\varphi,\varepsilon) = \frac{96}{\sqrt{\pi}}\chi\,\varphi^2
			\end{equation}
			\begin{equation}
				\overline{T}(\varphi, \varepsilon) = \frac{T}{\sigma^2\dot{\gamma}^2} = \frac{F_2(\varphi,\varepsilon)}{(1-\varepsilon^2)F_5
				(\varphi,\varepsilon)}
			\end{equation}
			\begin{equation}
				\overline{P}(\varphi,\varepsilon) = \frac{P}{\sigma^2\dot{\gamma}^2} =
				\frac{F_1(\varphi,\varepsilon)F_2(\varphi,\varepsilon)}{\varphi(1-\varepsilon^2)F_5(\varphi,\varepsilon)}
			\end{equation}
			\begin{equation}
				\overline{\tau}(\varphi,\varepsilon) = \frac{F_2(\varphi,\varepsilon)^{3/2}}{2\varphi(1-\varepsilon^2)^{1/2}F_5(\varphi,\varepsilon)^{1/2}}
			\end{equation}
			\begin{equation}
				\mu(\varphi,\varepsilon) = \frac{\tau}{P} = \frac{\overline{\tau}(\varphi,\varepsilon)}{\overline{P}(\varphi,\varepsilon)}
			\end{equation}
		\end{subequations}

\begin{figure}
			\begin{center}
				\includegraphics[scale=0.55]{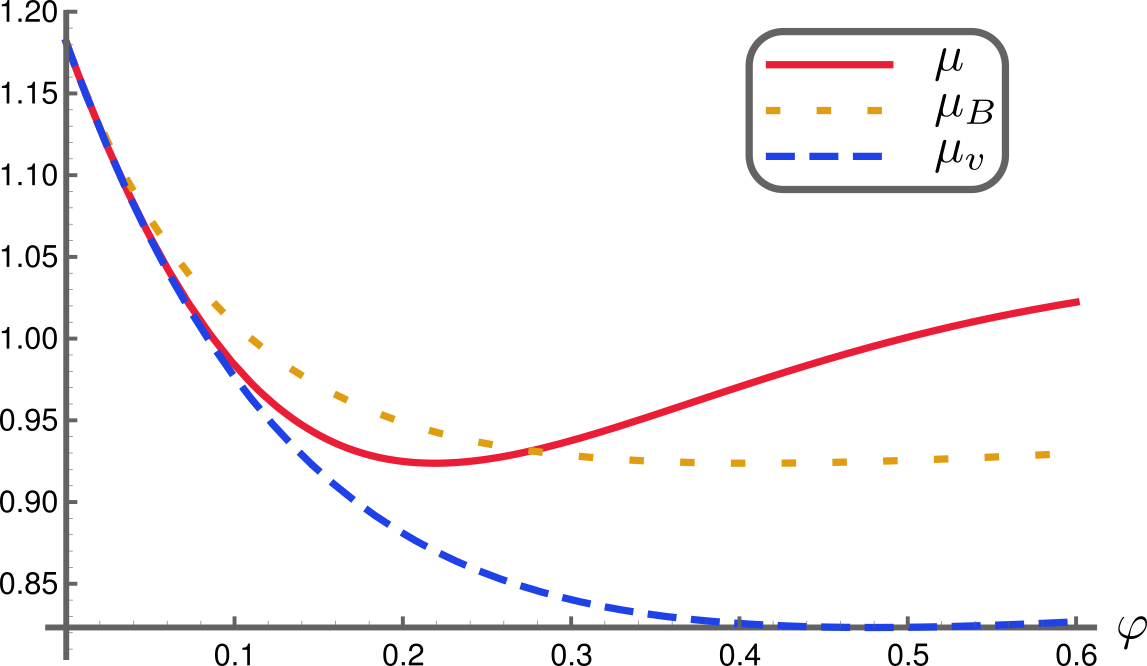}
			\end{center}
			\caption{Evolution of the effective friction coefficient in the dilute elastic model as a function of the packing fraction.
			The full line corresponds to the full expression of $\mu$, the dotted line to its expression in the Boltzmann limit, and
			the dashed line to its virial expansion at order 2.}
			\label{figLun1}
		\end{figure}

		Results of these computations can be visualised on figure \ref{figLun1}.
		On this figure, we first represented the full expression of the effective friction coefficient $\mu$.
		Two regimes can clearly be distinguished~: a dilute regime in which $\mu$ is a decreasing function of $\varphi$, and a denser
		regime in which it increases with $\varphi$.
		For this graph, we used the Carnaham-Starling value of $\chi$ for numerical evaluations, and the restitution coefficient is taken to 
		be $\varepsilon=0.85$, a rather good approximation of its value for a typical granular system (glass beads for example).

		In order to get a better understanding of this behavior, we examined two truncations of the expression of $\mu$~: First, we considered the Boltzmann limit
		in which the pair correlation function at contact does not appear in the Boltzmann equation (this term has to be added in a purely \textit{ad hoc}
		fashion); which corresponds to setting $\chi=1$.
		The comparison with the full value of $\mu$ shows that only in the very dilute limit the two curves meet each other.
		The evolution of the Boltzmann friction coefficient $\mu_B$ shows no convincing correlation to the two regimes identified before.
		Thus, the change of regime cannot be attributed to a pure effect of the value of the pair correlation function at contact.

		Then, we represented the value of the virial expansion of the effective friction coefficient truncated at order 2 in $\varphi$.
		This curve shows a much more convincing profile correspondence to the full $\mu$ in the dilute limit (let us stress that this is by no mean
		trivial, the virial expansion guaranteeing only an equality in the $\varphi\rightarrow0$ limit, which could lead to huge differences in
		the $\varphi\neq0$ regime, as is the case for the virial expansion of the pair correlation function of hard spheres near its contact value
		for example \cite{Coquand20b}).

\begin{figure}
			\begin{center}
				\includegraphics[scale=0.55]{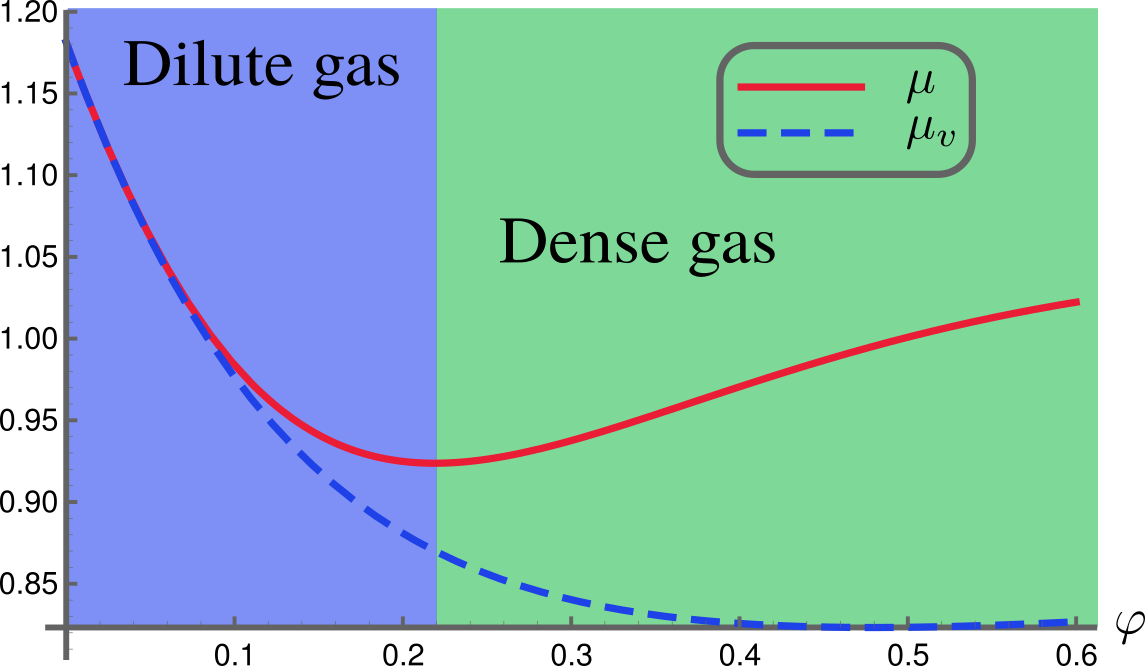}
			\end{center}
			\caption{Evolution of the effective friction coefficient in the dilute elastic model as a function of the packing fraction.
			The full line corresponds to the full expression of $\mu$, and the dashed line to its virial expansion at order 2.
			Two regimes of evolution are identified.}
			\label{figLun2}
		\end{figure}

		Another interesting point is that the virial expansion seems to correspond quite well to the first part of the evolution of the effective
		friction coefficient.
		This gives a further reason to label the low $\varphi$ regime the \textit{dilute gas} regime, as the physics in this regime is controlled by
		its virial limit.
		Around $\varphi\simeq0.2$, other effects, beyond the virial expansion, come into play, and the evolution of the effective friction coefficient
		changes.
		By contrast to the first regime, this regime is called the \textit{dense gas} regime (see figure \ref{figLun2}).

		Remark that $\mu$ remains an increasing function of $\varphi$ up to $\varphi=0.6$, which contradicts the $\mu(\mathcal{I})$ law
		that predicts an opposite evolution.
		This is a clear criterion to establish that the dilute elastic model does not capture the physics of the granular liquid.
		As we are going to see in the following, this feature is shared by all the other models based on the Boltzmann equation.

		From now on, in order to discriminate between the different models, the effective friction coefficient of the dilute elastic model will be called
		$\mu_{de}$.

	\subsection{The Sheared Enskog Expansion}

		For detailed computations in the sheared Enskog expansion model, the reader is referred to \cite{Montanero99}.

		In this model, the contributions to the macroscopic observables are split according to their origin~: for example, the pressure is
		written $P^{tot} = P^k + P^c$, where the kinetic pressure $P^k$ contains contributions from the left hand side of Eq.~(\ref{eqBo}),
		whereas the collisional pressure $P^c$ contains contributions from the modified collision integral (right hand side of Eq.~(\ref{eqBo})).

		Then, the collision integral is projected onto an orthonormal basis of functions constructed from the triplet $\big\{1,\mathbf{v},v^2\big\}$.
		Finally, a dimensionless shear rate
		\begin{equation}
			a^* = \frac{5}{4}\frac{\dot\gamma}{\omega_c} = \frac{5}{4}\,\text{Pe}\,,
		\end{equation}
		(Pe being the dry granular Peclet number \cite{Kranz20}) is built and expanded as follows~:
		\begin{equation}
			a^* \underset{\varepsilon\rightarrow1}{=} \overline{a}_1\,(1-\varepsilon^2)^{1/2} + \overline{a}_3\,(1-\varepsilon^2)^{3/2}
			+ O\big((1-\varepsilon^2)^{5/2}\big)\,.
		\end{equation}
		Finally, this expansion is plugged into the expressions of the macroscopic observables to yield an expression valid in the elastic limit.
		Remark that, compared to the dilute elastic model, this construction preserves much more of the non perturbative $\varphi$ dependence of the
		various physical quantities; It is thus expected to be more precise in the dense gas regime.

		The final expressions are as follows \cite{Montanero99}~:
		\begin{subequations}
			\begin{equation}
				\upsilon(\varphi) = n(\varphi)\,\chi(\varphi)
			\end{equation}
			\begin{equation}
				A_{xy}^{(1)}(\varphi) = -\frac{2\pi}{15}\upsilon(\varphi) 
			\end{equation}
			\begin{equation}
				A_{xy}^{(3)}(\varphi) = -\frac{2\pi}{105}\upsilon(\varphi)\left[14\,\alpha_2(\varphi) + \frac{128\pi}{25}\upsilon^2(\varphi)\right]
			\end{equation}
			\begin{equation}
				A_{xx}^{(2)}(\varphi) = \frac{128\pi}{1575}\upsilon^2(\varphi)
			\end{equation}
			\begin{equation}
				\tau^{k,(1)}_{xy}(\varphi) = 1 + \frac{4\pi}{15}\upsilon(\varphi)
			\end{equation}
			\begin{equation}
				\tau^{c,(1)}_{xy}(\varphi) = \frac{4\pi}{15}\upsilon(\varphi)\left[1-2\,A_{xy}^{(1)}(\varphi) + \frac{16}{5}\upsilon(\varphi)\right]
			\end{equation}
			\begin{equation}
				\alpha_2(\varphi) = \frac{\tau^{k,(1)}_{xy}(\varphi) + \tau^{c,(1)}_{xy}(\varphi)}{2\,\omega_0}
			\end{equation}
			\begin{equation}
				\omega_0 = \frac{5}{8}
			\end{equation}
			\begin{equation}
				\tau_{xx}^{k,(2)}(\varphi) = \frac{4}{3}\left[1 + \frac{4\pi}{15}\upsilon(\varphi) + \frac{64\pi}{525}\upsilon^2(\varphi)\right]
			\end{equation}
			\begin{equation}
				\tau_{yy}^{k,(2)}(\varphi) =-\frac{2}{3}\left[1 + \frac{4\pi}{15}\upsilon(\varphi) - \frac{128\pi}{525}\upsilon^2(\varphi)\right]
			\end{equation}
			\begin{equation}
				\tau_{zz}^{k,(2)}(\varphi) =-\frac{2}{3}\left[1 + \frac{4\pi}{15}\upsilon(\varphi) + \frac{256\pi}{525}\upsilon^2(\varphi)\right]
			\end{equation}
			\begin{equation}
				\begin{split}
					\tau_{xy}^{k,(3)}(\varphi) = & -\frac{2}{3}\left[1 + \frac{68\pi}{75}\upsilon(\varphi) +\frac{128\pi}{75}\left(\frac{\pi}{5}
							-\frac{1}{7}\right)\upsilon^2(\varphi)\right. \\
								& \left. +\frac{256\pi^2}{1875}\left(\frac{\pi}{3} + \frac{13}{7}\right)\upsilon^3(\varphi)\right]
				\end{split}
			\end{equation}
			\begin{equation}
				\begin{split}
					\tau_{xx}^{c,(2)}(\varphi)= & \frac{16\pi}{45}\upsilon(\varphi)\left[\left(1 + \frac{36}{35}\upsilon(\varphi)\right)
						\left(1 - 2A_{xy}^{(1)}(\varphi)\right) \right. \\
								    & \left. + \frac{144\pi}{175}\upsilon^2(\varphi) + \frac{3}{2}A_{xx}^{(2)}(\varphi)\right]
								    +\frac{\pi}{3}\alpha_2(\varphi)\upsilon(\varphi)
				\end{split}
			\end{equation}
			\begin{equation}
				\begin{split}
					\tau_{yy}^{c,(2)}(\varphi)= & \frac{16\pi}{45}\upsilon(\varphi)\left[-\left(\frac{1}{2} - \frac{36}{35}\upsilon(\varphi)\right)
						\left(1 - 2A_{xy}^{(1)}(\varphi)\right) \right. \\
								    & \left. + \frac{144\pi}{175}\upsilon^2(\varphi) + \frac{3}{2}A_{xx}^{(2)}(\varphi)\right]
								    +\frac{\pi}{3}\alpha_2(\varphi)\upsilon(\varphi)
				\end{split}
			\end{equation}
			\begin{equation}
				\begin{split}
					\tau_{zz}^{c,(2)}(\varphi)= & \frac{16\pi}{45}\upsilon(\varphi)\left[-\left(\frac{1}{2} - \frac{12}{35}\upsilon(\varphi)\right)
						\left(1 - 2A_{xy}^{(1)}(\varphi)\right) \right. \\
								    & \left. + \frac{48\pi}{175}\upsilon^2(\varphi) - 3\,A_{xx}^{(2)}(\varphi)\right]
								    +\frac{\pi}{3}\alpha_2(\varphi)\upsilon(\varphi)
				\end{split}
			\end{equation}
			\begin{equation}
				\begin{split}
					\tau_{xy}^{c,(3)}(\varphi)= & -\frac{16\pi}{45}\upsilon(\varphi)\left[\left(\frac{1}{2} - \frac{6}{35}\upsilon(\varphi)
						A_{xy}^{(1)}(\varphi) + \frac{1}{35}\upsilon(\varphi)\right) \right. \\
								    & \times \left(1 - 2A_{xy}^{(1)}(\varphi)\right) - \frac{256\pi}{875}\upsilon^3(\varphi) \\
								    & \left. - \frac{3}{2}\left(1 + \frac{32}{35}\upsilon(\varphi)\right)\,A_{xx}^{(2)}(\varphi)
								    + \frac{3}{2}A_{xy}^{(3)}(\varphi)\right] \\
								    & +\frac{1}{2}\alpha_2(\varphi)\tau_{xy}^{c,(1)}(\varphi)
				\end{split}
			\end{equation}
			\begin{equation}
				\begin{split}
					\alpha_4(\varphi) = -\frac{1}{2\omega_0} & \left[ \tau_{xy}^{k,(3)}(\varphi) + \tau_{xy}^{c,(3)}(\varphi) \right.\\
					& + \alpha_2(\varphi)\big(\alpha_2(\varphi)\omega_0 + 2\,\omega_2(\varphi)\big)\Big]
				\end{split}
			\end{equation}
			\begin{equation}
				\omega_2(\varphi) = \frac{3}{32} + \frac{7\pi}{30}\upsilon(\varphi) + \pi\left(\frac{8}{25} + \frac{\pi}{18}\right)\upsilon^2(\varphi)
			\end{equation}
			\begin{equation}
				\overline{a}_3(\varphi) = \frac{\overline{a}_1(\varphi)^5}{2\,\omega_0}\left[\frac{\omega_2(\varphi)}{\overline{a}_1(\varphi)^2}
				- \tau_{xy}^{k,(3)}(\varphi) - \tau_{xy}^{c,(3)}(\varphi)\right]
			\end{equation}
			\begin{equation}
				\overline{a}_1(\varphi) = \frac{1}{\sqrt{-\alpha_2(\varphi)}}
			\end{equation}
			\begin{equation}
			\label{eqastar}
				a^*(\varphi,\varepsilon) = \overline{a}_1\,(1-\varepsilon^2)^{1/2} + \overline{a}_3\,(1-\varepsilon^2)^{3/2}
			\end{equation}
			\begin{equation}
				\begin{split}
					\tau_{xy}(\varphi,\varepsilon) & = \tau(\varphi,\varepsilon) \\
								       & = a^*(\varphi,\varepsilon)\big(\tau_{xy}^{k,(1)}(\varphi) + \tau_{xy}^{c,(1)}(\varphi)\big) \\
								       & + a^*(\varphi,\varepsilon)\left(\overline{a}_1(\varphi)\sqrt{1-\varepsilon^2}\right)^2
									       \big(\tau_{xy}^{k,(3)}(\varphi) + \tau_{xy}^{c,(3)}(\varphi)\big)
				\end{split}
			\end{equation}
			\begin{equation}
			\label{eqP0}
				P_0(\varphi) = 1 + \frac{2\pi}{3}\upsilon(\varphi)
			\end{equation}
			\begin{equation}
			\label{eqPtot}
				\begin{split}
					&P^{tot}(\varphi,\varepsilon)  = P_0(\varphi) + \frac{a^*(\varphi,\varepsilon)}{3}^2\big[  \tau_{xx}^{k,(2)}(\varphi,\varepsilon)
						+ \tau_{xx}^{c,(2)}(\varphi,\varepsilon) \\
																& +\tau_{yy}^{k,(2)}(\varphi,\varepsilon)
						+ \tau_{yy}^{c,(2)}(\varphi,\varepsilon) + \tau_{zz}^{k,(2)}(\varphi,\varepsilon)
					+ \tau_{zz}^{c,(2)}(\varphi,\varepsilon)\big]
				\end{split}
			\end{equation}
			\begin{equation}
				\mu(\varphi,\varepsilon) = \frac{\tau_{xy}(\varphi,\varepsilon)}{P^{tot}(\varphi,\varepsilon)}
			\end{equation}
		\end{subequations}

		Remark that Eq.~(\ref{eqastar}) means that in practice, the expansion in powers of $(1-\varepsilon)$ is used truncated at the next-to-leading order.
		The bare pressure $P_0(\varphi)$ is the value of the pressure of a dilute unsheared gas of elastic spheres.

		A particularly interesting feature of this formalism is its ability to separate the contributions of the kinetic and collisional contributions
		to the rheological observables.

		\begin{figure}
			\begin{center}
				\includegraphics[scale=0.55]{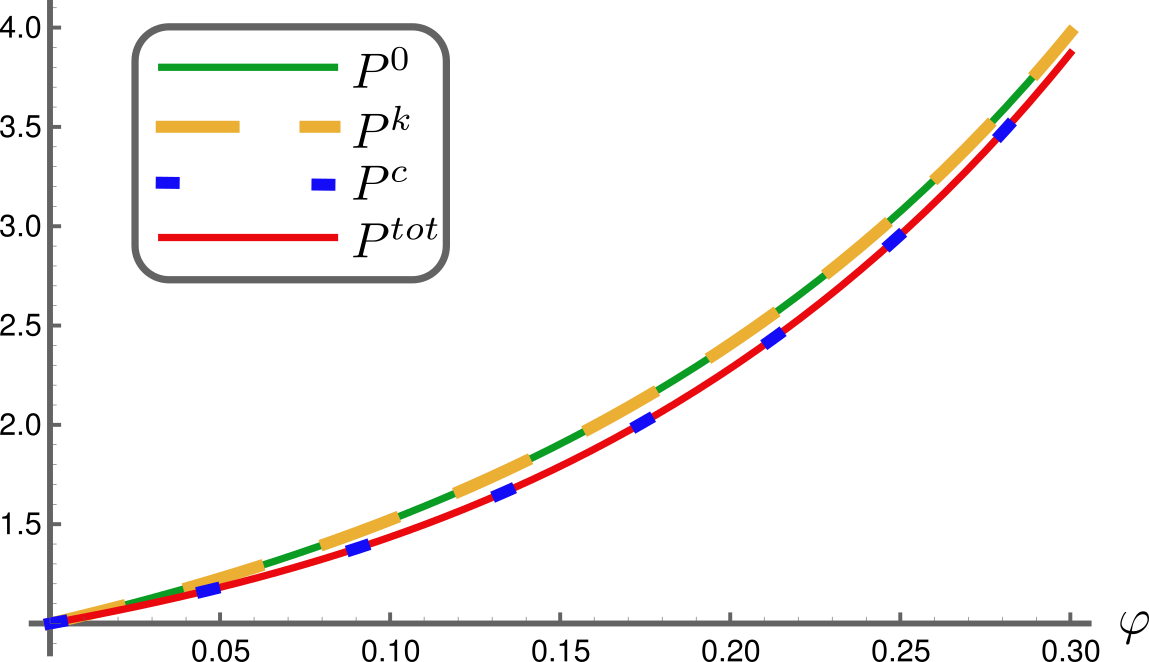}
			\end{center}
			\caption{Evolution of the pressure in the granular gas as a function of the packing fraction for a restitution coefficient 
			$\varepsilon=0.85$.
			We plotted separately the unsheared dilute elastic pressure $P_0(\varphi)$, the kinetic contribution $P^k(\varphi)$,
			the collisional pressure $P^c(\varphi)$ and the total pressure $P^{tot}(\varphi)$.}
			\label{figPS1}
		\end{figure}

		Let us first discuss the pressure.
		On Fig.~\ref{figPS1}, we compare the evolution of the bare pressure $P_0$, the kinetic pressure $P^{k}$ obtained by putting to 0 all the
		collisional contributions in Eq.~(\ref{eqPtot}), the collisional pressure obtained by putting to 0 all the kinetic contributions in
		Eq.~(\ref{eqPtot}), and the total pressure $P^{tot}$.
		First, we can see that all four values of the pressure are very similar~: the total pressure in the granular gas is almost unaffected
		by the contributions of the inelastic terms.
		A more thorough study shows that $P^k(\varphi)-P_0(\varphi)\simeq0$ on the whole range $\varphi\in[0;0.6]$.
		The collisional contribution leads to a slight decrease of the value of the pressure compared to its bare value.
		Note that this conclusion is at odds with the observation that in the liquid regime, as long as the Bagnold equation holds,
		the pressure gets a significant correction due to the forced advection of particles \cite{Coquand20f}.
		This should not come as a surprise as we are going to see that, similarly to the dilute elastic model, the sheared Enskog expansion
		does not describe well the rheology of the granular liquid.

		\begin{figure}
			\begin{center}
				\includegraphics[scale=0.55]{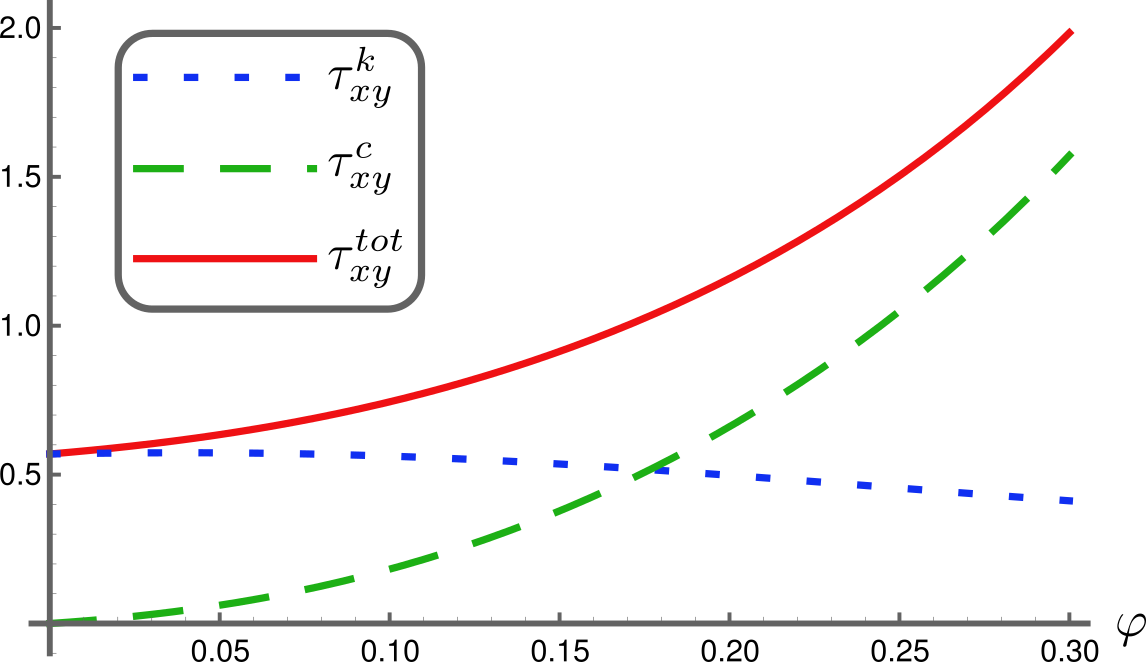}
			\end{center}
			\caption{Evolution of the shear stress in the granular gas as a function of the packing fraction for a restitution coefficient 
			$\varepsilon=0.85$.
			We plotted separately the kinetic contribution to the shear stress $\tau^k_{xy}(\varphi)$,
			the collisional contribution $\tau^c_{xy}(\varphi)$ and the total shear stress $\tau_{xy}^{tot}(\varphi)$.}
			\label{figPS2}
		\end{figure}

		Then, on Fig.~\ref{figPS2}, we represented the evolution of the different contributions to the shear stress $\tau$.
		Here, we can see a clear crossover between a very dilute regime where the value of the shear stress is dominated by the
		kinetic term, and a denser regime where the collisional part takes over.
		This is very interesting insofar as it means that the denser regime corresponds to corrections brought by complex collision processes
		beyond the binary collision terms.

		\begin{figure}
			\begin{center}
				\includegraphics[scale=0.55]{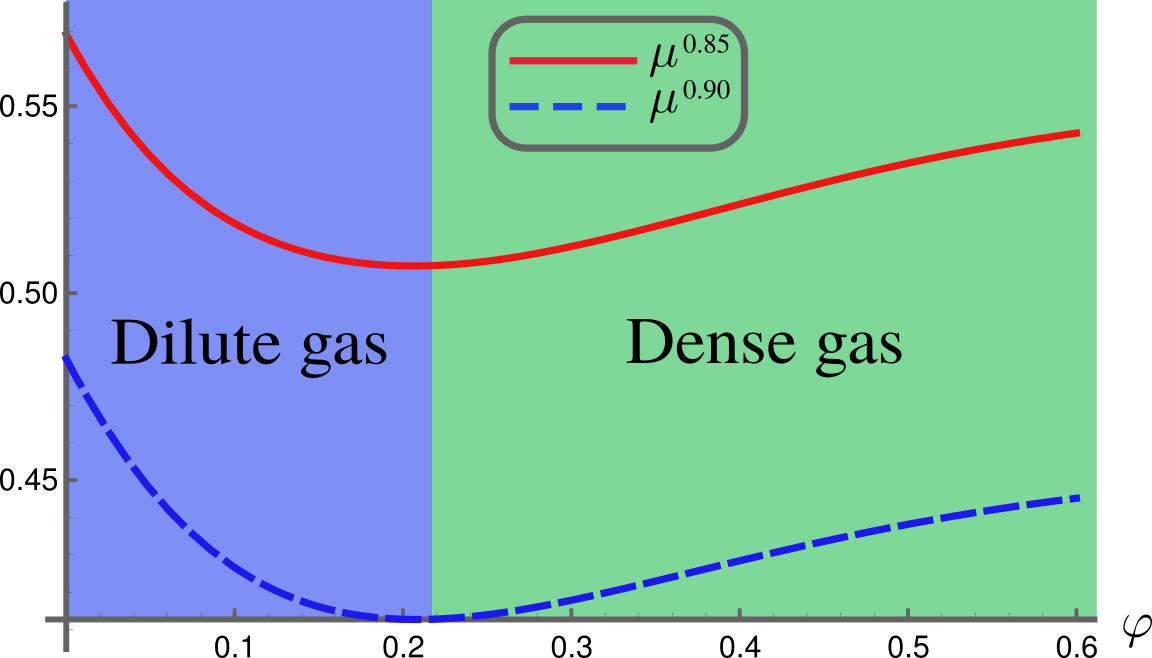}
			\end{center}
			\caption{Evolution of the effective friction coefficient in the granular gas as a function of the packing fraction for restitution coefficients 
			$\varepsilon=0.85$ and $\varepsilon=0.90$.}
			\label{figMuS1}
		\end{figure}

		Let us now have a look at the evolution of the effective friction coefficient, depicted on Fig.~\ref{figMuS1}.
		We displayed its value for two values of the effective friction coefficient.
		First, we can observe that the previous transition between a dilute and a dense regime is recovered in this model as well.
		Furthermore, the value of the crossover packing fraction, which is very close for the two curves (but this is also due to the
		fact that the chosen values of $\varepsilon$ are quite close to each other) is compatible with the crossover value identified on
		the shear stress evolution.

		\begin{figure}
			\begin{center}
				\includegraphics[scale=0.55]{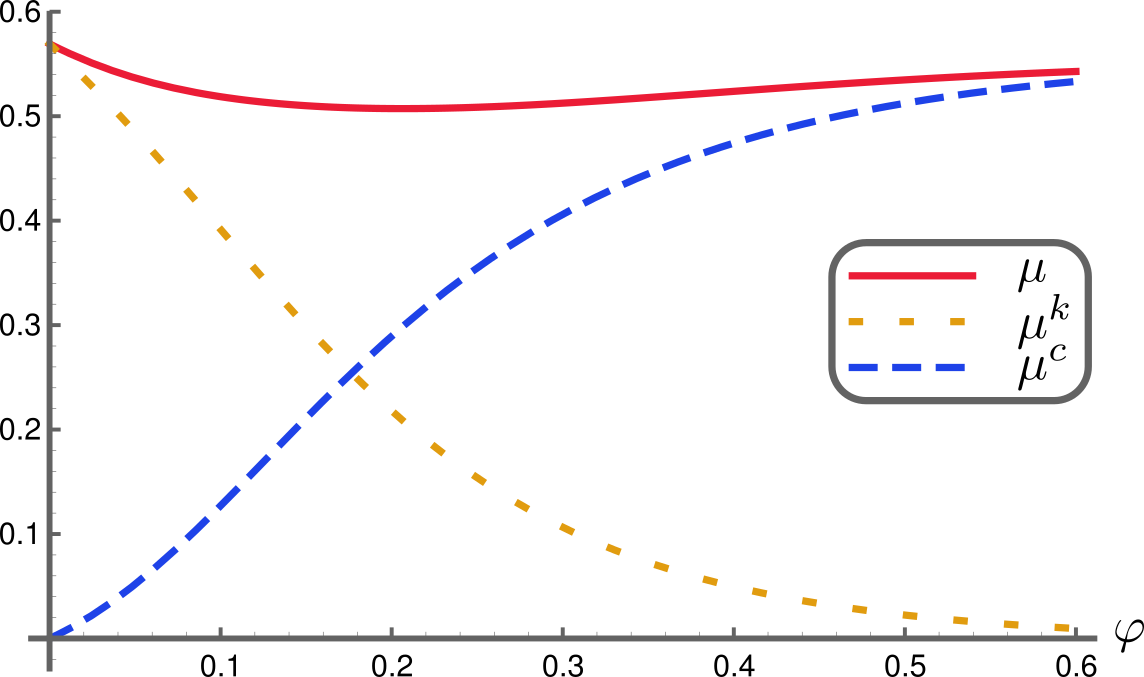}
			\end{center}
			\caption{Evolution of the effective friction coefficient in the granular gas as a function of the packing fraction for a restitution coefficient 
			$\varepsilon=0.85$.
			We plotted separately the kinetic contribution $\mu^k(\varphi)$,
			the collisional contribution $\mu^c(\varphi)$ and the total effective friction coefficient $\mu(\varphi)$.}
			\label{figMuS2}
		\end{figure}

		To verify this hypothesis, we displayed on Fig.~\ref{figMuS2} the evolution of $\mu$ as well as its different contributions.
		Given that the pressure varies very little depending on whether the collisional, or kinetic, contributions are taken into account,
		we kept $P^{tot}$ in the denominator, and changed only the value of the shear stress in the numerator.
		As expected, $\mu$ also shows a crossover between a kinetic and a collisional regime, with a crossover packing fraction ${\varphi\lesssim0.20}$.
		Interestingly, the two contributions appear to have opposite monotonycities, which reinforces our interpretation of their impact on the
		evolution of $\mu$.

		Finally, note that the value of $\mu$ changes quite significantly when $\varepsilon$ is varied, at least at the scale of the changes
		of $\mu$ as a function of $\varphi$.
		The qualitative shape of the curve, however, remains unchanged; In particular, the evolution at high $\varphi$s contradicts the
		$\mu(\mathcal{I})$ law, which shows that the sheared Enskog expansion model breaks down around $\varphi\simeq0.40$.

		In the following, the value of $\mu$ in the sheared Enskog expansion model will be denoted as $\mu_{sEe}$.

	\subsection{Inelastic Maxwell Models}

		For the third category of models, the reader is referred to \cite{Gonzalez19}, although it should be noted that the version of the model
		we present differs slightly for that of the original paper.

		The main idea behind this class of models is to replace the Enskog collision operator equation (\ref{eqJE}) by~:
		\begin{equation}
			\begin{split}
				\mathcal{J}_{IMM}\big[\mathbf{r},\mathbf{v}\big|f\big] &= \frac{\nu_M(\mathbf{r},t)}{4\pi\,n}\int d\mathbf{v}_2\int d\Omega \\
										       & \Big[\varepsilon^{-2}f^{(1)}(\mathbf{r},\mathbf{v}',t)f^{(1)}
											       (\mathbf{r}-\sigma,\mathbf{v}_2,t) \\
										       & -f^{(1)}(\mathbf{r},\mathbf{v},t)f^{(1)}(\mathbf{r}+\sigma,\mathbf{v}_2,t)\Big]\,,
			\end{split}
		\end{equation}
		where $\nu_M(\mathbf{r},t)$ is a phenomenological parameter that must be adjusted in order for the model to be solved.
		The main gain in this procedure is that the collision integral is greatly simplified.
		The quality of the model, however, depends on how well $\nu_M$ is adjusted.

		It is at this level that we depart from the original paper.
		Indeed, in the latter, the collision frequency is adjusted to the lowest order value of the pair correlation function at contact $\chi=1$;
		Yet, we have established above that this approximation is not adapted to our problem.
		Furthermore, the computation in \cite{Gonzalez19} takes for the bare pressure its lowest order value $P_0(\varphi)=1$, which is
		also not suitable for our purpose, which is why we replaced it by the elastic value of the hard sphere (dimensionless) pressure~:
		\begin{equation}
			P_0(\varphi) = 1 + 4\,\varphi\,\chi(\varphi)
		\end{equation}
		(it can be checked that this corresponds to the value chosen in the sheared Enskog expansion model Eq.~(\ref{eqP0})).

		All in all, the governing equations for the inelastic Maxwell model under those considerations are~:
		\begin{subequations}
			\begin{equation}
				\nu_0(\varphi) = \frac{4}{5}\omega_c(\varphi)
			\end{equation}
			\begin{equation}
				\zeta(\varphi,\varepsilon) = \frac{5}{12}(1-\varepsilon^2)\nu_0(\varphi)
			\end{equation}
			\begin{equation}
				\nu_{0|2}(\varphi,\varepsilon) = \zeta(\varphi,\varepsilon) + \frac{(1+\varepsilon)^2}{10}\nu_M(\varphi,\varepsilon)
			\end{equation}
			\begin{equation}
				\tau_{yy}^{IMM}(\varphi,\varepsilon) = \frac{\big(\nu_{0|2}(\varphi,\varepsilon)-\zeta(\varphi,\varepsilon)\big)}{\nu_{0|2}(\varphi,\varepsilon)}
				P_0(\varphi)
			\end{equation}
			\begin{equation}
				\tau_{xx}^{IMM}(\varphi,\varepsilon) = 3P_0(\varphi) - 2\,\tau_{yy}(\varphi,\varepsilon)
			\end{equation}
			\begin{equation}
				\overline{\dot{\gamma}}(\varphi,\varepsilon) = \frac{\sqrt{3\zeta(\varphi,\varepsilon)}\big(10\,\zeta(\varphi,\varepsilon) + 
				(1+\varepsilon)^2\nu_M(\varphi,\varepsilon)\big)}{2\sqrt{5}(1+\varepsilon)\sqrt{\nu_M(\varphi,\varepsilon)\,P_0(\varphi)}}
			\end{equation}
			\begin{equation}
				\tau_{xy}^{IMM}(\varphi,\varepsilon) = \frac{\overline{\dot{\gamma}}(\varphi,\varepsilon)}{\nu_{0|2}(\varphi,\varepsilon)}
				\tau_{yy}^{IMM}(\varphi,\varepsilon)
			\end{equation}
			\begin{equation}
				\mu_{IMM}(\varphi,\varepsilon) = \frac{3\,\tau_{xy}^{IMM}(\varphi,\varepsilon)}{\tau_{xx}^{IMM}(\varphi,\varepsilon) + 2\,\tau_{yy}^{IMM}
				(\varphi,\varepsilon)}
			\end{equation}
		\end{subequations}
		and the value of $\nu_M$ is adjusted so that $\tau_{xy}$ is consistent with its value in the sheared Enskog expansion in the elastic limit~:
		\begin{widetext}
			\begin{equation}
				\nu_M(\varphi,\varepsilon) = \frac{125\,\pi(1-\varphi)^3(1+\varphi+\varphi^2-\varphi^3)}{4\big(25\,\pi-70\,\pi\,\varphi
					+768\,\varphi^2 +159\,\pi\,\varphi^2 -768\,\varphi^3 - 204\,\pi\,\varphi^3 + 192\,\varphi^4 + 191\,\pi\,\varphi^4
				-110\,\pi\,\varphi^5 + 25\,\pi\,\varphi^6\big)}\,\nu_0(\varphi)
			\end{equation}
		\end{widetext}

\begin{figure}
			\begin{center}
				\includegraphics[scale=0.55]{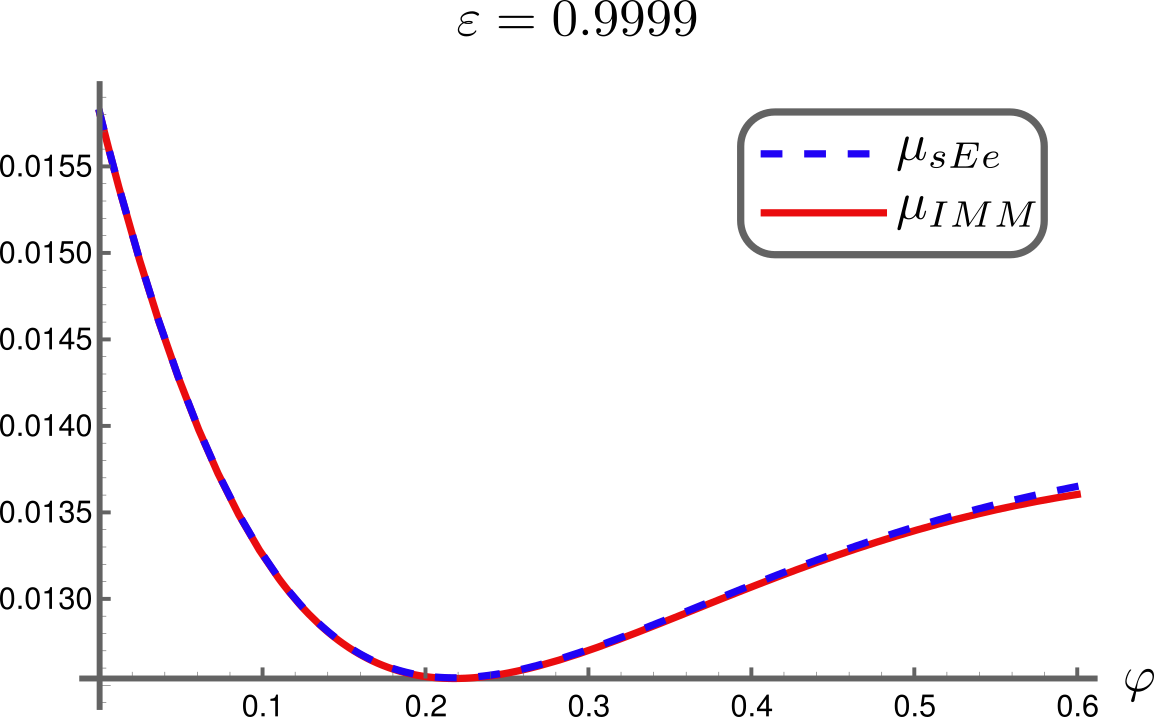}
			\end{center}
			\caption{Comparison between the effective friction coefficient of the sheared Enskog expansion model and the
			inelastic Maxwell model for $\varepsilon=0.9999$.}
			\label{figIMMMu1}
		\end{figure}

\begin{figure}
			\begin{center}
				\includegraphics[scale=0.55]{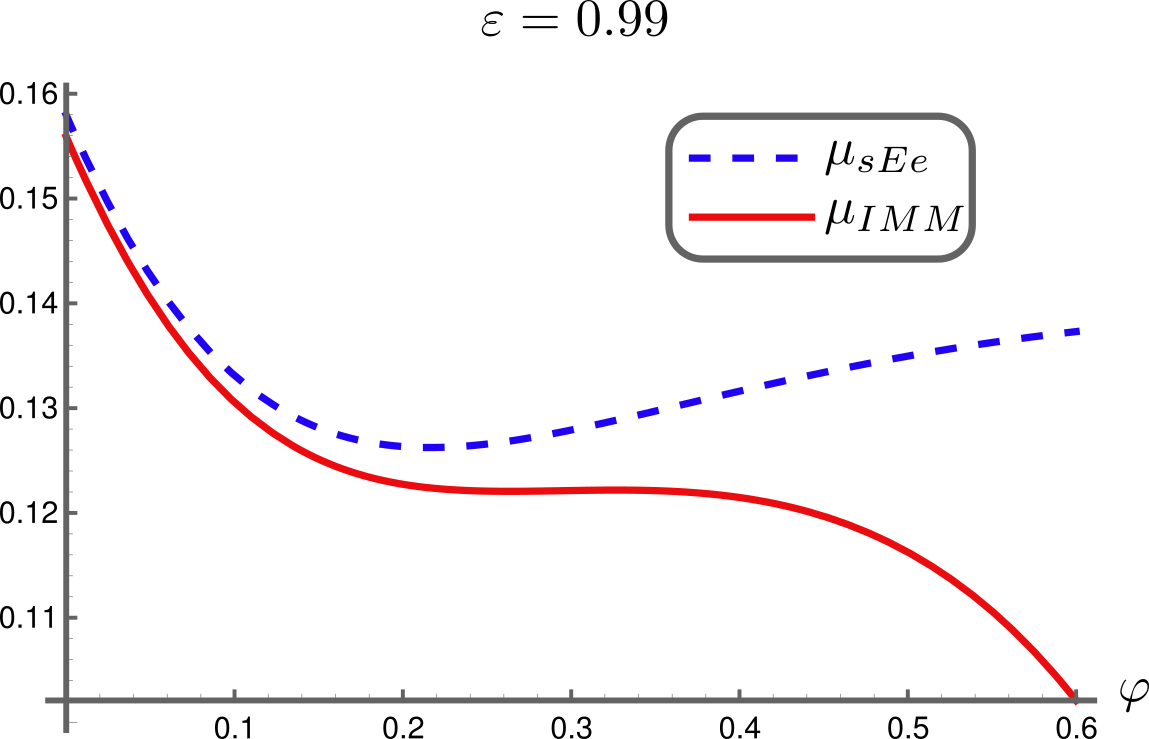}
			\end{center}
			\caption{Comparison between the effective friction coefficient of the sheared Enskog expansion model and the
			inelastic Maxwell model for $\varepsilon=0.99$.}
			\label{figIMMMu2}
		\end{figure}

\begin{figure}
			\begin{center}
				\includegraphics[scale=0.55]{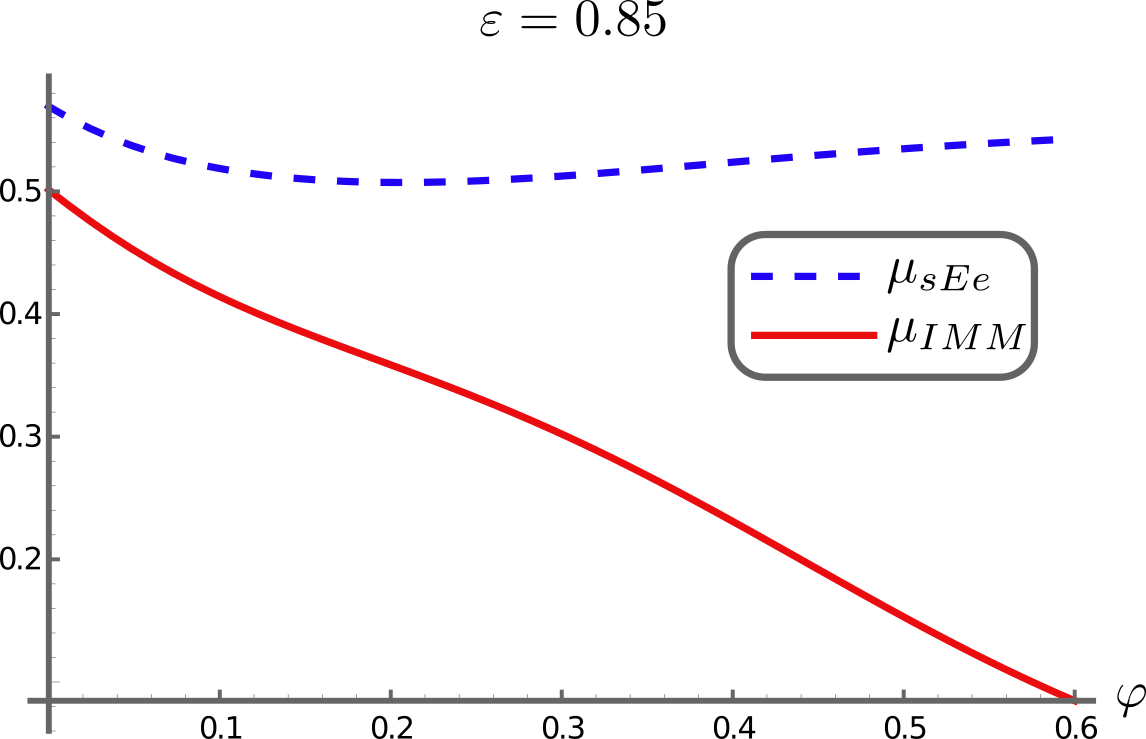}
			\end{center}
			\caption{Comparison between the effective friction coefficient of the sheared Enskog expansion model and the
			inelastic Maxwell model for $\varepsilon=0.85$.}
			\label{figIMMMu3}
		\end{figure}

		The results of this model are shown on Figs.~\ref{figIMMMu1}, \ref{figIMMMu2}, and \ref{figIMMMu3}.
		Consistently with the input we gave to the model, the sheared Enskog expansion is recovered in the elastic limit.
		However, $\varepsilon$ must be extremely close to 1 for the two models to yield compatible results.
		Worse, even at $\varepsilon-1=O\big(10^{-2}\big)$, the change in monotonicity is lost in the inelastic Maxwell model.
		Note that this observation is not a global statement on the performance of such models to describe properties of granular gases in general;
		\cite{Gonzalez19} and the references therein largely show the usefulness of inelastic Maxwell models.
		However, they seem not to be fit to an accurate description of the evolution of the effective friction coefficient outside of the elastic limit.

		One obvious source of precision loss in the model is the use of the bare pressure $P_0(\varphi)$ in place of the real expression of the pressure
		(dependent on $\varphi$ and $\varepsilon$).
		Unfortunately, this value is not known.
		That being said, we can still learn a bit more about the behavior of inelastic Maxwell models by studying the case where the pressure
		is given by its value in the sheared Enskog expansion (which is only an approximation of the real value of the pressure in the sheared
		granular gas).

\begin{figure}
			\begin{center}
				\includegraphics[scale=0.55]{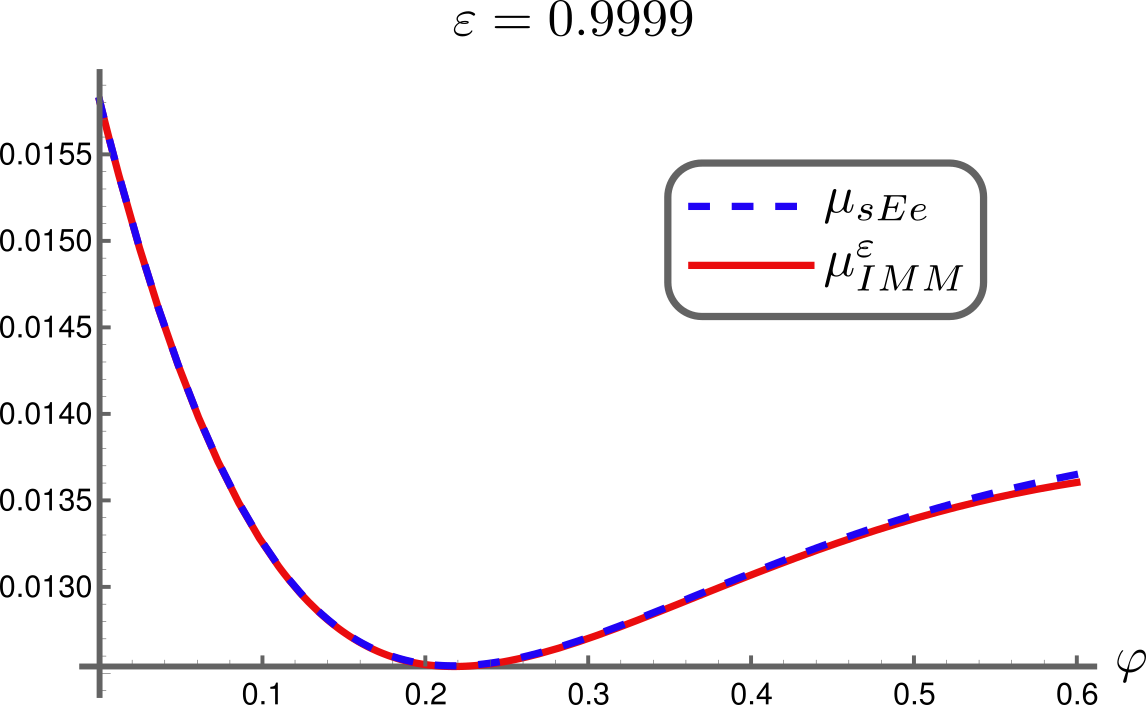}
			\end{center}
			\caption{Comparison between the effective friction coefficient of the sheared Enskog expansion model and the
			inelastic Maxwell model matched on the $\varepsilon$ expansion of the pressure for $\varepsilon=0.9999$.}
			\label{figIMMMue1}
		\end{figure}

\begin{figure}
			\begin{center}
				\includegraphics[scale=0.55]{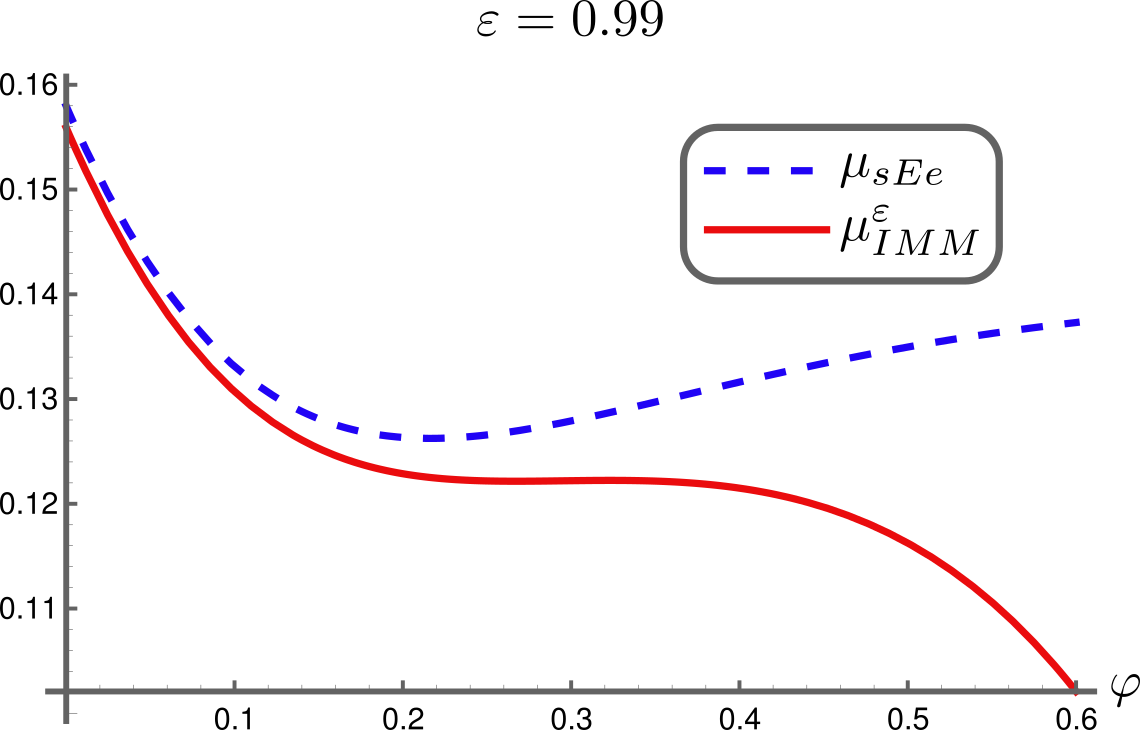}
			\end{center}
			\caption{Comparison between the effective friction coefficient of the sheared Enskog expansion model and the
			inelastic Maxwell model matched on the $\varepsilon$ expansion of the pressure for $\varepsilon=0.99$.}
			\label{figIMMMue2}
		\end{figure}

\begin{figure}
			\begin{center}
				\includegraphics[scale=0.55]{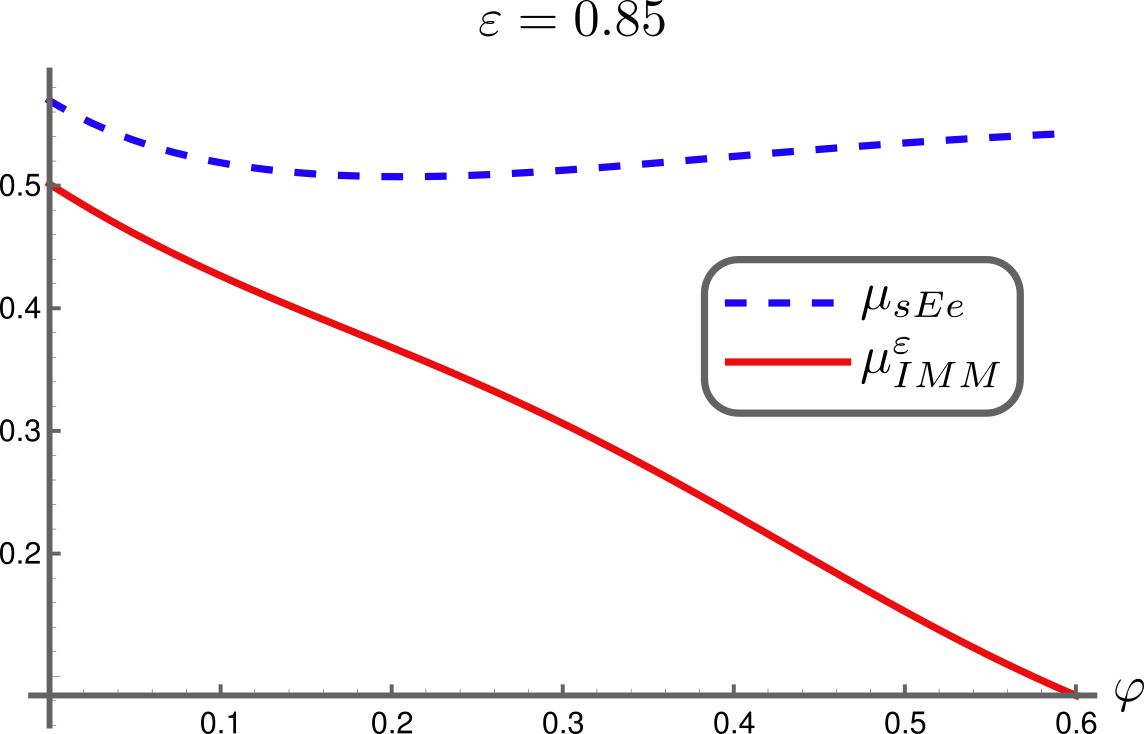}
			\end{center}
			\caption{Comparison between the effective friction coefficient of the sheared Enskog expansion model and the
			inelastic Maxwell model matched on the $\varepsilon$ expansion of the pressure for $\varepsilon=0.85$.}
			\label{figIMMMue3}
		\end{figure}

		We performed two expansions.
		In the first one, the pressure is truncated to order $O\big((\varepsilon-1)^2\big)$, keeping non-perturbative information on its $\varphi$
		dependence.
		The evolution of the corresponding effective friction coefficient, denoted $\mu_{IMM}^{\varepsilon}$ is represented on Figs.~\ref{figIMMMue1},
		\ref{figIMMMue2} and \ref{figIMMMue3}.

		Despite an enhanced content in the variations of the pressure with $\varphi$ (and crucially in the variation of the latter dependence with
		$\varepsilon$), there is little qualitative difference compared to the case where the bare pressure is used, even in the high $\varphi$
		region.
		Hence, the subleading $\varphi$ dependence of the pressure does not help to improve the quality of the inelastic Maxwell model,
		even in the high $\varphi$ region.

\begin{figure}
			\begin{center}
				\includegraphics[scale=0.55]{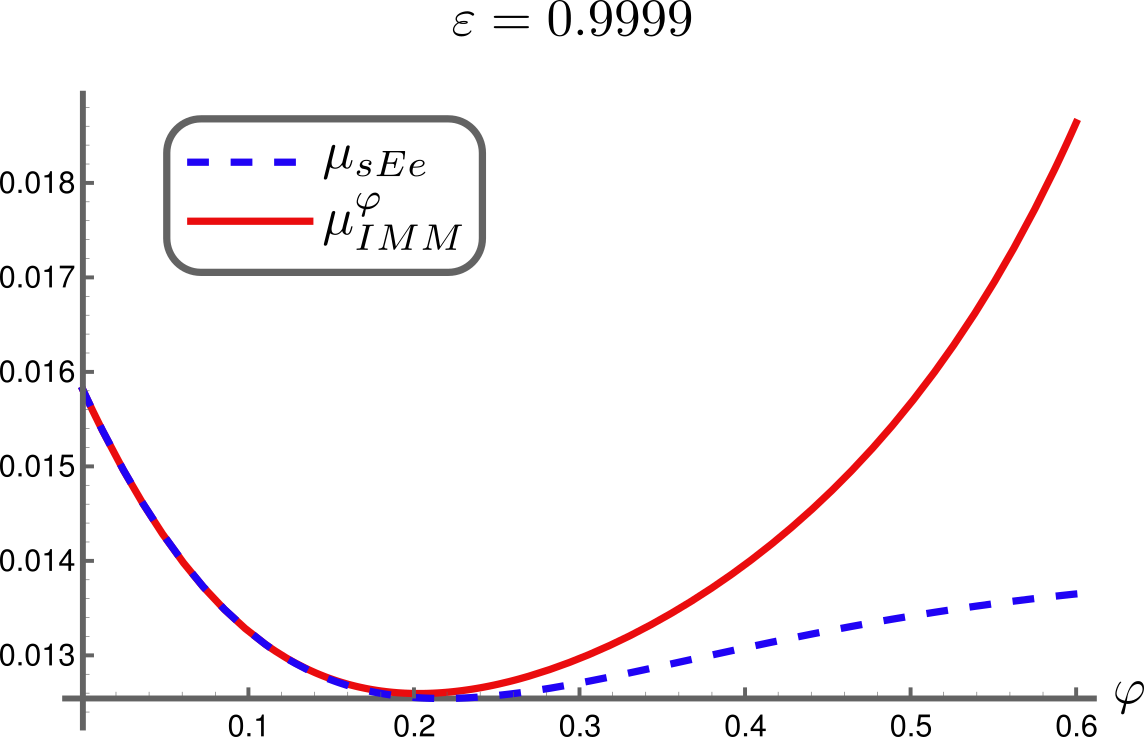}
			\end{center}
			\caption{Comparison between the effective friction coefficient of the sheared Enskog expansion model and the
			inelastic Maxwell model matched on the $\varphi$ expansion of the pressure for $\varepsilon=0.9999$.}
			\label{figIMMMuv1}
		\end{figure}

\begin{figure}
			\begin{center}
				\includegraphics[scale=0.55]{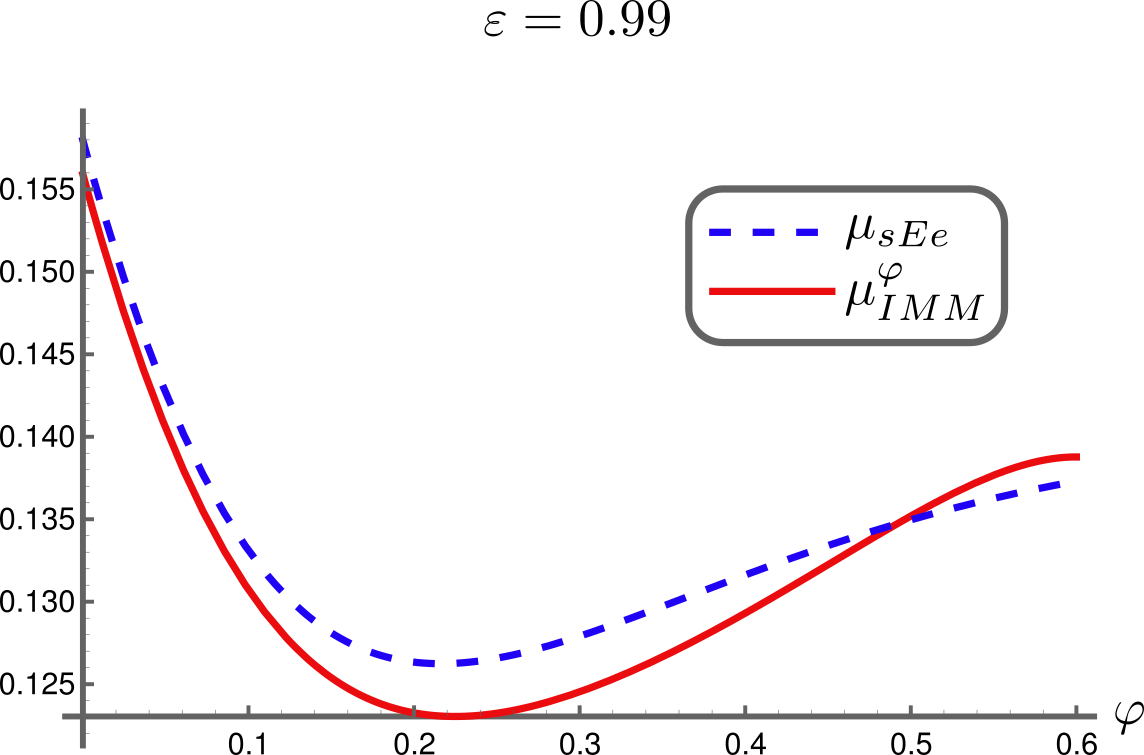}
			\end{center}
			\caption{Comparison between the effective friction coefficient of the sheared Enskog expansion model and the
			inelastic Maxwell model matched on the $\varphi$ expansion of the pressure for $\varepsilon=0.99$.}
			\label{figIMMMuv2}
		\end{figure}

\begin{figure}
			\begin{center}
				\includegraphics[scale=0.55]{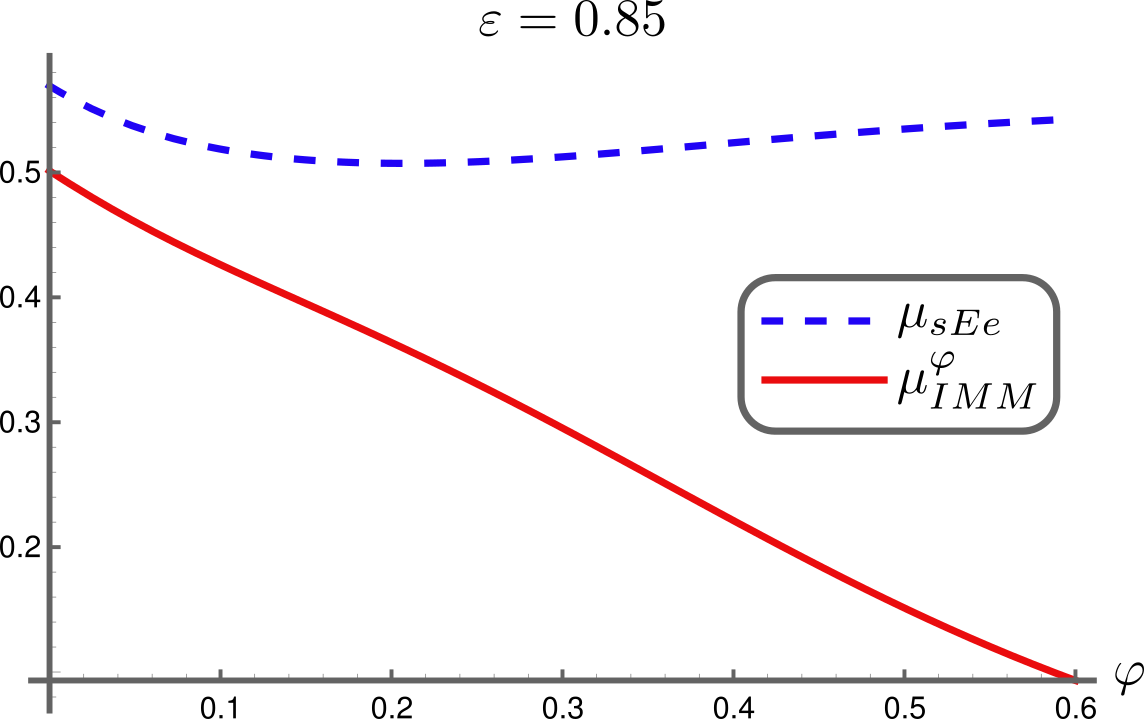}
			\end{center}
			\caption{Comparison between the effective friction coefficient of the sheared Enskog expansion model and the
			inelastic Maxwell model matched on the $\varphi$ expansion of the pressure for $\varepsilon=0.85$.}
			\label{figIMMMuv3}
		\end{figure}

		For the second expansion, we kept the non-perturbative $\varepsilon$ dependence, and performed a virial expansion of the sheared
		Enskog expansion model pressure at order $O\big(\varphi^4\big)$.
		The results are displayed on Figs.~\ref{figIMMMuv1}, \ref{figIMMMuv2} and \ref{figIMMMuv3}.

		Here, the qualitative pictures changes quite a lot.
		First, the agreement with the sheared Enskog expansion model is still guaranteed in the dilute elastic limit, but the effect
		of the virial expansion is clearly visible on Fig.~\ref{figIMMMuv1}, where even in the nearly elastic case, the high $\varphi$
		sector of the curve is wrong.
		This is a classical problem encountered when extrapolating a perturbative expansion away from the range where the expansion parameter is small.

		The second case, however, displayed on Fig.~\ref{figIMMMuv2}, shows a good qualitative agreement between the inelastic Maxwell model and
		the sheared Enskog expansion model, with the presence of the inversion of monotonicity, absent both from the case of the
		$\varepsilon-1$ expansion, and the case of the bare pressure.
		This shows that the non-perturbative content of the pressure dependence in $\varepsilon$ is a necessary feature to be able to recover
		a physics consistent with the sheared Enskog expansion and the dilute elastic model.

		Finally, Fig.~\ref{figIMMMuv3} shows that this agreement breaks down as the inelasticity of the granular particles is increased.
		This may be a sign that further $\varepsilon$ dependence, present only at higher orders in the virial expansion are needed to properly
		describe the evolution of the effective friction coefficient at smaller $\varepsilon$s.

		Let us emphasize that there is no point in going always further in these expansions to recover more content.
		It is clear that if we give the full sheared Enskog expansion model as an input, we should recover results somewhat in agreement
		with this model.
		Our approach here aims at discussing which part of the truncation used in the construction of the inelastic Maxwell model may be responsible
		for its inability to properly describe the evolution of the effective friction coefficient.

		We also want to make clear that the above discussion is by no way a proof that inelastic Maxwell models are not able to capture the
		correct rheology for dense granular gases.
		The only thing we argued is that the matchings we used for the parameter $\nu_M$ are not sufficient to reach this goal, but this leaves
		plenty of room for other kind of truncations to lead to more precise results.
		The reason we limit ourselves to the cases discussed above is that this work is particularly devoted to discussing the virial, and expansions
		around the elastic limit, that are the more commonly used approximations used to describe the rheology of granular gases (the closest
		known case being elastic hard spheres at equilibrium).

		Finally, let us discuss briefly another category of related kinetic models, the Bhatnagar-Gross-Krook (BGK) class, also presented in \cite{Gonzalez19}.
		For these models, the collision integral is replaced with the following phenomenological expression~:
		\begin{equation}
			\begin{split}
				\mathcal{J}_{BGK}\big[\mathbf{r},\mathbf{v}\big|f\big] &= \psi(\varphi,\varepsilon)\,\nu_0(\varphi)\big(f_{MB}(\mathbf{r},\mathbf{v},t)
										       -f^{(1)}(\mathbf{r},\mathbf{v},t)\big) \\
										       &+ \frac{\zeta(\varphi,\varepsilon)}{2}
										       \frac{\partial}{\partial\mathbf{V}}\cdot\mathbf{V}f^{(1)}(\mathbf{r},\mathbf{v},t)\,,
			\end{split}
		\end{equation}
		where $\psi$ is an adjustable parameter and $f_{MB}$ is the Maxwell-Boltzmann distribution function.
		In this case, it is possible to map exactly the solution of this model to an inelastic Maxwell model, the problem reduces to a polynomial
		expression of order two in $\psi$, which solutions $\psi_1$ and $\psi_2$ are not reproduced here due to their length.
		The elastic expansion yields $\displaystyle \psi_1\underset{\varepsilon\rightarrow1}{=}1 + O\big(\varepsilon-1\big)$, and
		$\displaystyle \psi_2\underset{\varepsilon\rightarrow1}{=}O\big((\varepsilon-1)^2\big)$, which enables to choose the appropriate solution.

		In the case of the study of the continuous/discontinuous shear thickening presented in \cite{Gonzalez19}, this change of perspective 
		allows to get more information on the content of the one particle distribution function $f^{(1)}$.
		Here however, as we focus on the rheological observables, it merely reproduces the results of the inelastic Maxwell models,
		and will therefore not be discussed further.

	\subsection{Discussion}

\begin{figure}
			\begin{center}
				\includegraphics[scale=0.55]{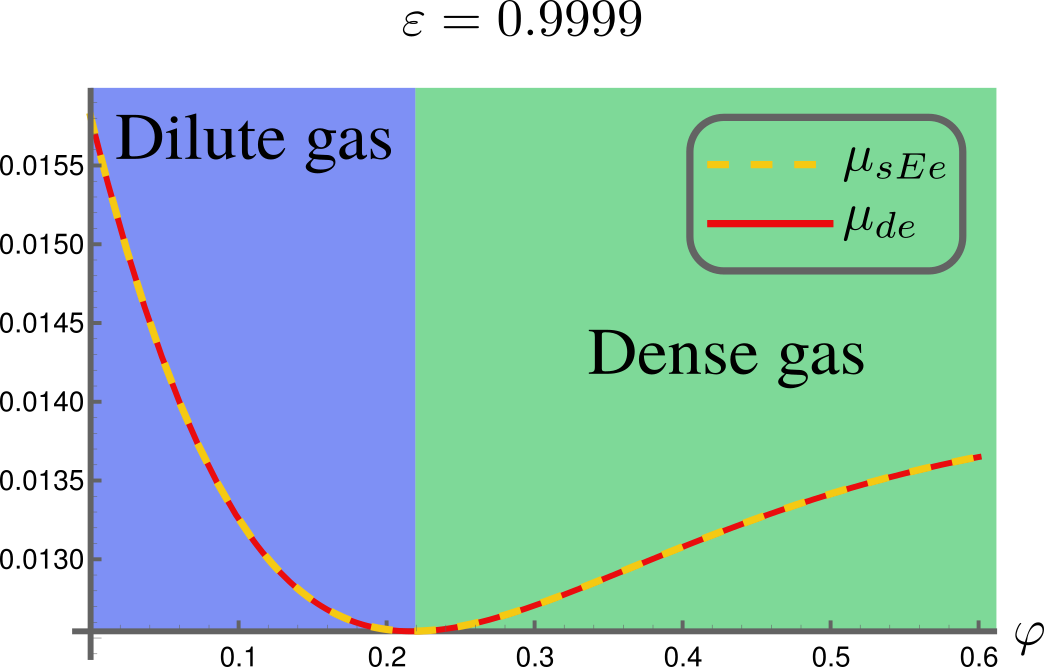}
			\end{center}
			\caption{Comparison between the effective friction coefficient of the sheared Enskog expansion model and the
			dilute elastic model for $\varepsilon=0.9999$.}
			\label{figMuMu1}
		\end{figure}

\begin{figure}
			\begin{center}
				\includegraphics[scale=0.55]{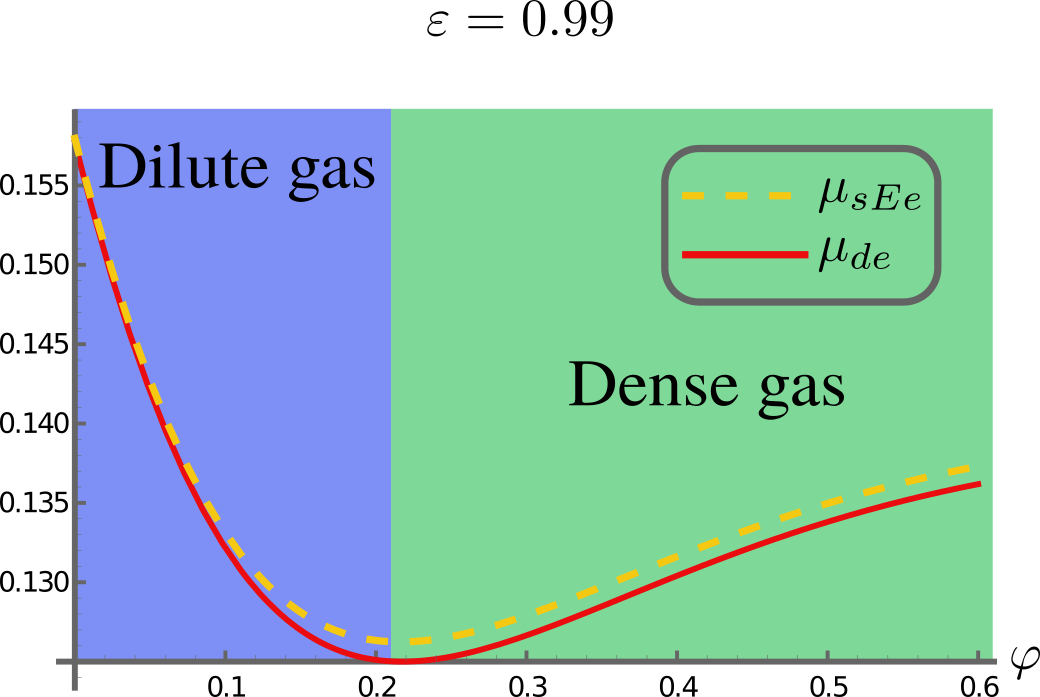}
			\end{center}
			\caption{Comparison between the effective friction coefficient of the sheared Enskog expansion model and the
			dilute elastic model for $\varepsilon=0.99$.}
			\label{figMuMu2}
		\end{figure}

\begin{figure}
			\begin{center}
				\includegraphics[scale=0.55]{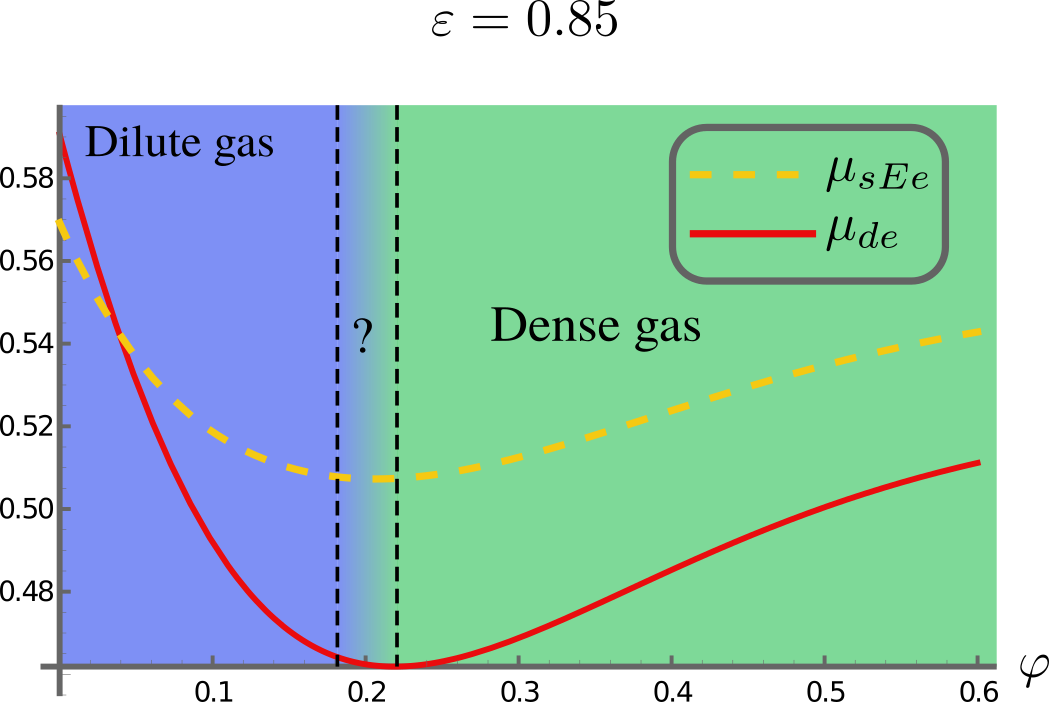}
			\end{center}
			\caption{Comparison between the effective friction coefficient of the sheared Enskog expansion model and the
			dilute elastic model for $\varepsilon=0.85$. The two black dashed lines indicate the approximate location
			of the transition between the two gaseous regimes for both model.}
			\label{figMuMu3}
		\end{figure}

		To conclude, the dilute elastic and sheared Enskog expansion models are going to be compared (the inelastic Maxwell models are not included
		due to their poor qualitative performance discussed above).
		The results of the comparisons for various values of $\varepsilon$ are shown on Figs.~\ref{figMuMu1}, \ref{figMuMu2} and \ref{figMuMu3}.

		First, there is a remarkable agreement between the two models at the qualitative level.
		As a check for the consistency of our construction, note that both models converge towards each other in the dilute elastic limit as shown
		on Fig.~\ref{figMuMu1}.
		Fig.~\ref{figMuMu2} is also interesting from this perspective as it shows a very good agreement at low $\varphi$, which quality
		decreases as one gets further away from the dilute limit.

		Another important point to mention is that this good agreement is only qualitative in nature, as evidenced by Fig.~\ref{figMuMu3}.
		On the latter, one can observe that already for a typical value of $\varepsilon$ (we did not really reach a strong inelasticity regime),
		the curves are clearly different from one another.
		Moreover, the value of the packing fraction at the transition between the dilute and the dense regime becomes blurry, as its value
		differs from one model to the other (see Fig.~\ref{figMuMu3}).
		Let us not overstress these differences, however, that remain mild at the scale of the values taken by $\mu$.
		The main point here is that, even for typical values of $\varepsilon$, both models are clearly distinguishable to the naked eye,
		which should be kept in mind when analysing the values predicted by one of such models for a given realistic system (the precision
		of these approaches is still a bit rough as far as $\mu$ is concerned).

		Finally, note the strong chute of the value of $\mu$ as the system gets closer and closer to the elastic limits.
		The values displayed on Figs.~\ref{figMuMu1} and \ref{figMuMu2} are not typical of what is generally measured experimentally on granular systems.

		All in all, as expected, the different models converge towards each other in the dilute elastic limit, but only in this case.
		They nonetheless make emerge a clear picture of the rheological behavior of a typical granular gas, with a dilute and a dense regimes,
		which physical interpretations have been presented before.
		Based on the way approximations are implemented in the two models, it is expected that the sheared Enskog expansion yields more precise results
		than the dilute elastic model, which is only valid in the limit $R\ll1$.
		However, a discussion of the above results at the light of experimental data would be very interesting.
		There is, to the best of our knowledge, no extensive study of the evolution of the effective friction coefficient of granular gases
		with the packing packing fraction (this observable has been mostly discussed in the case of the liquid, because, as we indicated before,
		it follows a strong universal law in this regime).
		Although measuring shear stress in the dilute regime may be a challenge, we believe that at least the increase of $\mu$ with $\varphi$
		in the dense gas regime should be measurable, with a reasonable degree of difficulty.

\section{The Rheology of Granular Liquids}

	Before going into the details of the GITT model, let us briefly recall the main properties of granular liquids.
	In most of the literature, granular liquids are separated from granular solids by their ability to flow, and from granular
	gases by the ruling of the rheology by the $\mu(\mathcal{I})$ law \cite{GDR04,Andreotti13}.

	However, the picture is a little bit more subtle. As a matter of fact, in the solid limit, corresponding to the $\mathcal{I}\rightarrow0$ limit,
	$\mu$ follows some non trivial power law $\displaystyle\mu \underset{\mathcal{I}\rightarrow0}{=} \mu_0 + A\,\mathcal{I}^\alpha$, where the
	exponent $\alpha$ depends on the microscopic friction coefficient between granular particles \cite{DeGiuli15,DeGiuli16,DeGiuli17a} (note the use
	of a constant $\mu_0$ which is a priori different from the $\mathcal{I}\rightarrow0$ limit of the $\mu(\mathcal{I})$ law, named $\mu_1$).
	More precisely, it is estimated that for $\mathcal{I}\lesssim 10^{-2.5}$, the granular material enters a friction dominated regime \cite{DeGiuli16},
	which rheology is clearly different from that of the liquid (in terms of packing fraction, this corresponds to $\varphi\simeq0.6$).
	For example, even for frictionless materials, the value of $\alpha$ is not an integer, which contradicts the prediction of the $\mu(\mathcal{I})$
	law.

	In the following, we identify the granular liquid regime as the regime in which the rheology is described by the $\mu(\mathcal{I})$ law.
	We will show for the first time, to the best of our knowledge, how this regime transitions to the gaseous regime at low packing fractions
	($\varphi\lesssim0.40$), and ask the reader to keep in mind that the following description does not holds up to the solid state, since the
	friction dominated regime will not be discussed.
	A last important remark is that in the granular liquid state, the $\mu(\mathcal{I})$ law holds irrespective of the fact that the particles are
	frictionless or frictional \cite{DaCruz05}.
	In particular, it is wrong to identify $\mu_1$ or $\mu_2$ in Eq.~(\ref{eqMuI}) with any form of microscopic friction coefficient between the granular particles.

	\subsection{Some Reminders of Mode-Coupling Theory}

		\subsubsection{Dynamics}

		The mode-coupling theory has originally been developed to describe the strong slow down of the dynamics in supercooled liquids \cite{Goetze08}.
		Indeed, it has been observed numerically \cite{Kob95} that the mean squared displacement $\left<r^2(t)\right>$ develops a plateau,
		separating the short time ballistic regime from the large time diffusive regime, that
		extends over several decades in the supercooled state.
		This plateau is interpreted as follows~: as the temperature is decreased, or the packing fraction is increased, particles tend to hinder
		the motion of their neighbors, this is called the \textit{cage effect}.

		For atomic, or molecular liquids, this cage effect is reduced by the possibility of thermoactived event where particles tunnel through
		the liquid, and increase its flowability.
		For granular liquids on the other hand, which are modelled as hard spheres, it is expected that the cage effect is strong, even at relatively
		high packing fractions.
		This is one of the reason of the success of MCT in the description of the rheology of granular liquids.

		The cage effect is a process by which the dynamics of the liquid is clearly strongly related to its internal structure.
		It is therefore rather natural to choose, as a central quantity in the description of the liquid's dynamics, a structure dependent quantity~:
		the first GITT equation is a MCT equation for the dynamical structure factor~:
		\begin{equation}
			\Phi_q(t) = N\frac{\left<\rho_q(t)\rho_{-q}(0)\right>}{S(q)}\,
		\end{equation}
		where $\rho$ is the density field, and $S(q)$ is the static structure factor (ensuring that $\Phi_q(0)=1$).
		The dynamics of $\Phi$ is described by a Mori-Zwanzig type equation, that we do not derive here (the reader is referred to \cite{Kranz20}),
		which writes~:
		\begin{subequations}
			\begin{equation}
				\begin{split}
					\ddot{\Phi}_{q}(t) &+ \nu_{q(t)}\,\dot{\Phi}_{q}(t) + q(t)^{2}C_{q(t)}^2\,\Phi_{q}(t) \\
							      & +q(t)^{2}C_{q(t)}^2\,\int_0^t d\tau\,m_{q}(t,\tau)\dot{\Phi}_{q}(\tau) = 0
				\end{split}
			\label{eqMCT}
			\end{equation}
			\begin{equation}
				\nu_q = \frac{1+\varepsilon}{3}\,\omega_c\,\Big[1 +3 j_0''(q\sigma)\Big]
			\end{equation}
			\begin{equation}
				j_0''(x) = \frac{d^2}{dx^2}\left[\frac{\sin(x)}{x}\right]
			\end{equation}
			\begin{equation}
				C_q^2 = \frac{T}{S(q)}\left[\frac{1 + \varepsilon}{2} + \frac{1 - \varepsilon}{2}S(q)\right]
			\end{equation}
		\end{subequations}
		and $m_q(t,\tau)$ is a memory kernel that contains two terms \cite{Alder70,Resibois75a,Resibois75b,Furtado76,Bosse78,Sjogren80,Kranz14}~:
		a first term $m_q^{(1)}$ corresponding to binary collisions, containing a Dirac-$\delta$ function which makes it independent of the previous history
		of the system, and a second term $m_q^{(2)}$ containing the physics of the dynamics slow down triggered by collective excitations that writes
		\cite{Kranz20}~:
		\begin{equation}
		\label{eqmMCT}
			\begin{split}
				m_q^{(2)}(t,\tau) &= A_{q(t)}(\varepsilon)\frac{S\big(q(t)\big)}{nq^2}\int\frac{d^3k}{(2\pi)^3} S\big(k(\tau)\big)
				S\big(p(\tau)\big) \\
				&\times\Big[(\hat{\mathbf{q}}.\mathbf{k})nc\big(k(t)\big) + (\hat{\mathbf{q}}.\mathbf{p})nc\big(p(t)\big)\Big]\\
				&\times\Big[(\hat{\mathbf{q}}.\mathbf{k})nc\big(k(\tau)\big) + (\hat{\mathbf{q}}.\mathbf{p})nc\big(p(\tau)\big)\Big] \\
				&\times\Phi_{k(\tau)}(t-\tau)\Phi_{p(\tau)}(t-\tau) \,,
			\end{split}
		\end{equation}
		where $\mathbf{q} = \mathbf{k} + \mathbf{p}$, the autocorrelation function $c(q)$ is given by the Ornstein-Zernicke equation~:
		\begin{equation}
			S(q) = \frac{1}{1 - n\,c(q)}\,,
		\end{equation}
		and the inelasticity damping factor is~:
		\begin{equation}
			A_q^{-1}(\varepsilon) = 1 + \frac{1-\varepsilon}{1 + \varepsilon}\,S(q)\,.
		\end{equation}
		Finally, in all these equations, $q(t)$ is the wave vector advected by the shear flow~:
		\begin{equation}
			q^2(t) = q^2\left(1 + \frac{\dot{\gamma}^2t^2}{3}\right)\,.
		\end{equation}

		In most of the recent literature about MCT, $m_q^{(1)}$ is omitted.
		This is because MCT is mostly used in the supercooled state of liquids, where this term is completely negligible (the dynamics is dominated
		by the collective excitations).
		We have explicitly verified in the case of GITT that the presence of this term does not change the numerical values for the rheological
		quantities in the liquid state that have been previously published in \cite{Coquand20f,Coquand20g,Coquand21}.
		When the packing fraction is decreased however, single collisions become the dominant mechanism, as discussed earlier.
		There should therefore be a crossover at moderate packing fractions, that needs both terms of the memory kernel to be described properly.
		We give more details on this later.

		\begin{figure}
			\begin{center}
				\includegraphics[scale = 0.55]{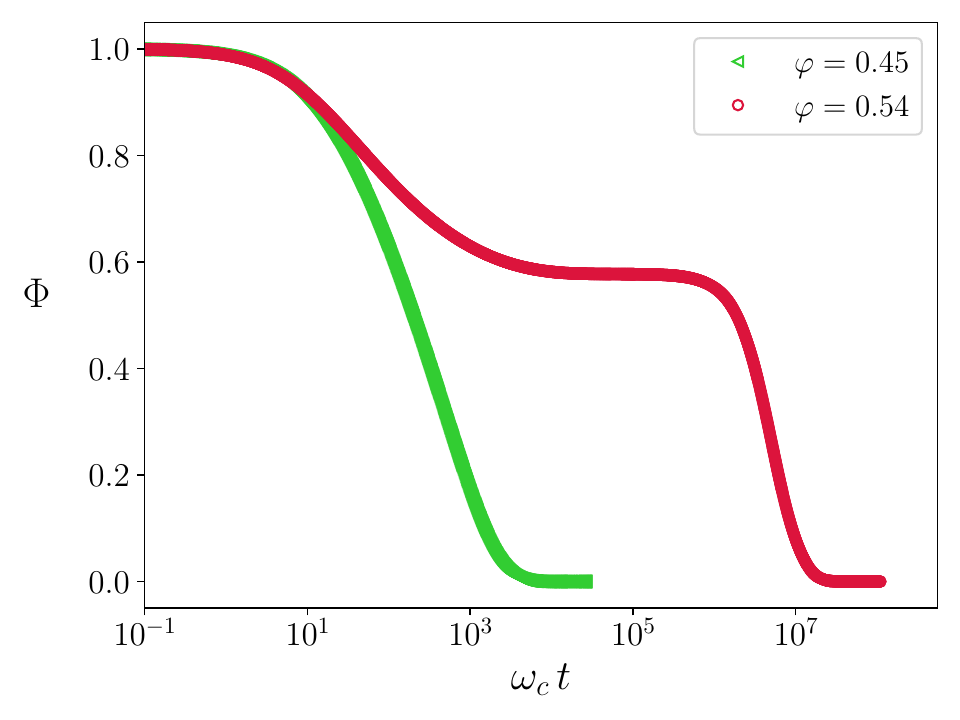}
			\end{center}
			\caption{Typical evolution of the dynamical structure factor with time as a function of the packing fraction.
			These curves correspond to $\varepsilon=0.85$ and Pe=$\dot{\gamma}/\omega_c=2\cdot10^{-7}$.}
			\label{figPhi}
		\end{figure}

		The typical solutions to Eq.~(\ref{eqMCT}) are displayed on Fig.~\ref{figPhi} as a function of time (the dependence on the wave number
		plays a subleading role in the rheology and will not be discussed further).
		We chose a low value of the Peclet number Pe (corresponding to a low shear rate) in order to accentuate the slow down of the dynamics.
		It appears that at low packing fractions, $\Phi$ shows a single step decay at $t\simeq\omega_c^{-1}$.
		This is the regime where collective excitations are weak, Eq.~(\ref{eqMCT}) is dominated by the first three terms, and $\Phi$ therefore
		decays exponentially.
		At higher packing fractions however, a plateau develops, that correspond to the plateau in the mean squared displacement discussed earlier,
		and which causes $\Phi$ to decay on a much larger time scale.
		Fig.~\ref{figPhi} corresponds to a low Pe; In the Bagnold regime, Pe reaches its maximum value Pe$^{\text{max}}$, and the length of the plateau
		is greatly reduced \cite{Coquand20g}.

		\subsubsection{Linking dynamics to the rheology~: ITT}

		In the MCT formalism, the rheological observables are defined at the \textit{microscopic scale} in a Liouville equation,
		dressed up by the interactions at all mesoscopic scales, yielding the measurable, \textit{macroscopic} observables
		written as statistical averages of the microscopic operators \cite{Kranz20}.
		For example, the measured shear stress $\tau_{xy}^{GITT}$ is related to the microscopic shear stress $\tau_{xy}^{mic}$ by~:
		\begin{equation}
			\tau_{xy}^{GITT} = \left<\tau_{xy}^{mic}\right>^{(\dot\gamma)}\,,
		\end{equation}
		where the $(\dot\gamma)$ exponent makes explicit the fact that the statistical ensemble corresponds to the out of equilibrium sheared fluid.

		The definition of the latter ensemble is very challenging with the tools of today's statistical physics.
		This is where the ITT approach comes into play.
		It was originally developed in the context of the study of the rheology of dense colloids \cite{Fuchs02,Fuchs09}.
		It consists in expressing the out of equilibrium statistical averages on the sheared system $\left<\cdot\right>^{(\dot\gamma)}$ as
		a function of statistical averages taken on a \textit{fictitious}, \textit{quiescent} unsheared fluid, which is therefore at equilibrium.
		These averages are denoted $\left<\cdot\right>_0$.
		The price to pay for such a replacement is the arising of an integral term that runs over all the history of the system's deformations \cite{Fuchs02}.

		In granular fluids, the situation is a bit more subtle since~: (i) granular fluids, even unsheared, can never be at equilibrium (except in
		a trivial state), and (ii) due to the dissipative character of collisions, a source of fluidization must be introduced in the quiescent fluid
		to maintain a flowing state.
		The subtleties of the implementation of the ITT formalism to granular fluids are discussed in further details in \cite{Kranz20}.
		For now, let us accept that the macroscopic observables can be computed in the granular fluid states thanks to this method.

		The results have been derived in \cite{Kranz20} for the shear stress, and \cite{Coquand20f} for the pressure.
		They can be expressed as~:
		\begin{subequations}
			\begin{equation}
			\label{eqTauGITT}
				\begin{split}
					\tau^{GITT}_{xy} &= \tau_{xy}^E \\
							 &+\frac{1}{60\pi^2}\int_0^{+\infty}dt\frac{1}{\sqrt{1 + \big(\dot{\gamma}t\big)^2/3}}\int_0^{+\infty}
					dk\,\mathcal{F}_1(k,t)
				\end{split}
			\end{equation}
			\begin{equation}
				\begin{split}
					P^{GITT}& = P_0^{GITT} \\
						&+ \frac{1}{36\pi^2}\int_0^{+\infty}dt\frac{\big(\dot{\gamma}t\big)}{\sqrt{1 + \big(\dot{\gamma}t\big)^2/3}}\int_0^{+\infty}
					dk\,\mathcal{F}_1(k,t) \\
						&+ \frac{1}{12\pi^2}\int_0^{+\infty}dt\frac{\big(\dot{\gamma}t\big)}{\sqrt{1 + \big(\dot{\gamma}t\big)^2/3}}\int_0^{+\infty}
					dk\,\mathcal{F}_2(k,t) \\
				\end{split}
			\end{equation}
			\begin{equation}
				\mathcal{F}_1(k,t) = -k^4\,\dot{\gamma}T\left(\frac{1 + \varepsilon}{2}\right)\Phi_{k(-t)}^2\frac{S'(k(-t))S'(k)}{S(k)^2}
			\end{equation}
			\begin{equation}
				\mathcal{F}_2(k,t) = -k^3\,\dot{\gamma}T\left(\frac{1 + \varepsilon}{2}\right)\Phi_{k(-t)}^2\frac{S'(k(-t))\big(S(k)^2-S(k)\big)}{S(k)^2}
			\end{equation}
			\begin{equation}
				S'(k) = \frac{dS}{dk}(k)
			\end{equation}
		\end{subequations}
		In the first equation, $\tau_{xy}^E$ is the contribution of $m_q^{(1)}$, which is matched to the elastic limit of the value of $\tau_{xy}$
		in the sheared Enskog expansion (note that no truncation is made on the remaining $\varphi$ content).
		This is done so that both approaches should be consistent in the dilute elastic limit.
		We have made explicit that the bare pressure $P_0^{GITT}$ is slightly different from the expression used above.
		As a matter of fact, in order to improve the precision of the computation in the liquid regime ($\varphi\geqslant0.40$), we used for the
		value of the pair correlation function at contact, the value derived by Clisby and Mc Coy \cite{Clisby06}, based on a resummation of the
		virial series \cite{Coquand20f}.
		In practice, this value is quasi-indistinguishable from that of Carnaham-Starling at lower packing fractions, so we can still use it
		here without problem of consistency (see \cite{Coquand20f} for a detailed discussion).

		\subsubsection{Energy conservation}

		In order to close our system of equations, we need a last one, which is the energy conservation equation~:
		\begin{equation}
		\label{eqPow}
			\mathcal{P}_D + \tau^{GITT}_{xy}\,\dot{\gamma} = n\,\Gamma_d\,\omega_c\,T \,,
		\end{equation}
		where $\mathcal{P}_D$ is a fluidization power.
		In the Bagnold regime, which is the main focus of this paper, $\mathcal{P}_D=0$.
		We put it in the equation to recall that, when working with the quiescent fictitious fluid in ITT, we have to add $\mathcal{P}_D\neq 0$
		to ensure that the granular temperature of the quiescent fluid is the same as that of the sheared one.
		For more details about this procedure, the reader is referred to \cite{Kranz20}.

		In the Bagnold regime, Eq.~(\ref{eqPow}) can be rewritten as~:
		\begin{equation}
		\label{eqPow2}
			\overline{\tau}_{xy}^{GITT}\,\text{Pe}^{\text{max}} = \Gamma_d \,,
		\end{equation}
		with $\overline{\tau} = \tau/nT$ and Pe$=\dot{\gamma}/\omega_c$.
		When $\mathcal{P}_D\neq0$ in Eq.~(\ref{eqPow}), a wide variety of shear rates $\dot{\gamma}$ can be reached.
		In the Bagnold regime however, we have established that $T\propto\dot{\gamma}^2$, thus, since $\omega_c\propto\sqrt{T}$,
		Pe becomes independent of the value of $\dot{\gamma}$.
		It reaches its highest value, hence its notation Pe$^{\text{max}}$.

		Although it may not be obvious at first glance, the balance equation (\ref{eqPow2}) is a self consistent one (the value of Pe$^{\text{max}}$
		depends on the value of $\overline{\tau}_{xy}^{GITT}$).
		In practice, it is solved as such, by a dichotomy algorithm.
		This is a major difference with the sheared Enskog expansion as here, we do not truncate the series as a function of the restitution coefficient
		$\varepsilon$; The value of Pe$^{\text{max}}$ is fully non-perturbative, both in $\varepsilon$ and in $\varphi$.
		As we will discuss later, this does not come without implications when comparing both approaches.

		\subsubsection{A time scale analysis of $\Phi$'s decay}

		The GITT set of equations presented above is much too complicated to be solved analytically.
		However, we showed in a previous study \cite{Coquand20g} that it is suited to the development of simple toy models that cast the
		behavior of rheological observables in terms of a simple time scale comparison problem.

		As for the other models, we cannot reproduce here the full derivation of the toy model, the interested reader is referred to \cite{Coquand20g}
		for details.
		The starting point is the remark that the decay of the $\Phi_q(t)$ function can occur only through two channels~: structural relaxations
		of the fluid's structure (with rate $\Gamma$), and advection (motion of particles imposed by the shear).
		Since the $q$-dependence of $\Phi_q(t)$ only yields subleading effects, we can proceed to the replacement~:
		\begin{equation}
			\Phi_q(t)\mapsto e^{-\Gamma t}e^{-\dot{\gamma}^2t^2/\gamma_c^2}
		\end{equation}
		in the GITT equations, which can then be solved analytically ($\gamma_c$ is a typical strain scale of the material).

		It has been established that the rheology of granular liquids, and the evolution of $\mu$ in particular, can then be explained in terms
		of competition of three time scales~: the structural relaxation time scale $t_\Gamma=1/\Gamma$, the advection time scale
		$t_\gamma=1/\dot{\gamma}$, and the free fall time scale $t_{ff}$ presented in the introduction.
		From these three time scales, two dimensionless numbers are built~: the Weissenberg number Wi$=t_\Gamma/t_\gamma$ and the
		inertial number $\mathcal{I}=t_{ff}/t_\gamma\propto$ Pe.
		The (unique) lowest order Padé approximant that captures convincingly the behavior of $\mu$ can be expressed in terms of these
		numbers as \cite{Coquand20g}~:
		\begin{equation}
		\label{eqMuGITT}
			\mu_{GITT}\big(\text{Wi},\mathcal{I}\big) = \frac{\mu_1}{1 + \text{Wi}_0/\text{Wi}} + \frac{\mu_2-\mu_1}{1 + \mathcal{I}_0/\mathcal{I}} \,,
		\end{equation}
		where Wi$_0$ is a constant.
		Note that this Padé structure is by no mean obvious, for example, in other flow geometries such as extensional flows, the lowest order
		Padé approximants involve more terms, which allows for a richer world of possibilities for the evolution of $\mu$ \cite{Coquand21}.

		In the Bagnold regime, the shear rate $\dot{\gamma}$ reaches its maximum value in units of the collision frequency $\omega_c$.
		Advection is thus the governing channel for the relaxation of $\Phi$~: the shear tends to force particles to move, and is thus able to break
		cages in regimes of high density where the equivalent unsheared system would be arrested.
		In other words, the structural relaxation time scale $t_\Gamma$ does not play any role in the rheology, it effectively decouples~:
		\begin{equation}
			t_\gamma\ll t_\Gamma\quad\Rightarrow\quad \text{Wi}\gg\text{Wi}_0\,,
		\end{equation}
		and the $\mu(\mathcal{I})$ law Eq.~(\ref{eqMuI}) is recovered.

		Let us emphasize that, although, consistently with dimensional analysis, $\mu$ in the Bagnold regime depends on only one dimensionless number ($\mathcal{I}$),
		its full expression Eq.~(\ref{eqMuGITT}) involves a third time scale, which is still indirectly present in the $\mu(\mathcal{I})$ formula
		in the $\mu_1$ coefficient.
		If, in addition to the shear, the granular medium were fluidized, $\mu_1$ would get screened accordingly (but in this case, the Bagnold
		equation does not hold anymore).

	\subsection{The effective friction coefficient in the GITT model}

\begin{figure}
			\begin{center}
				\includegraphics[scale=0.55]{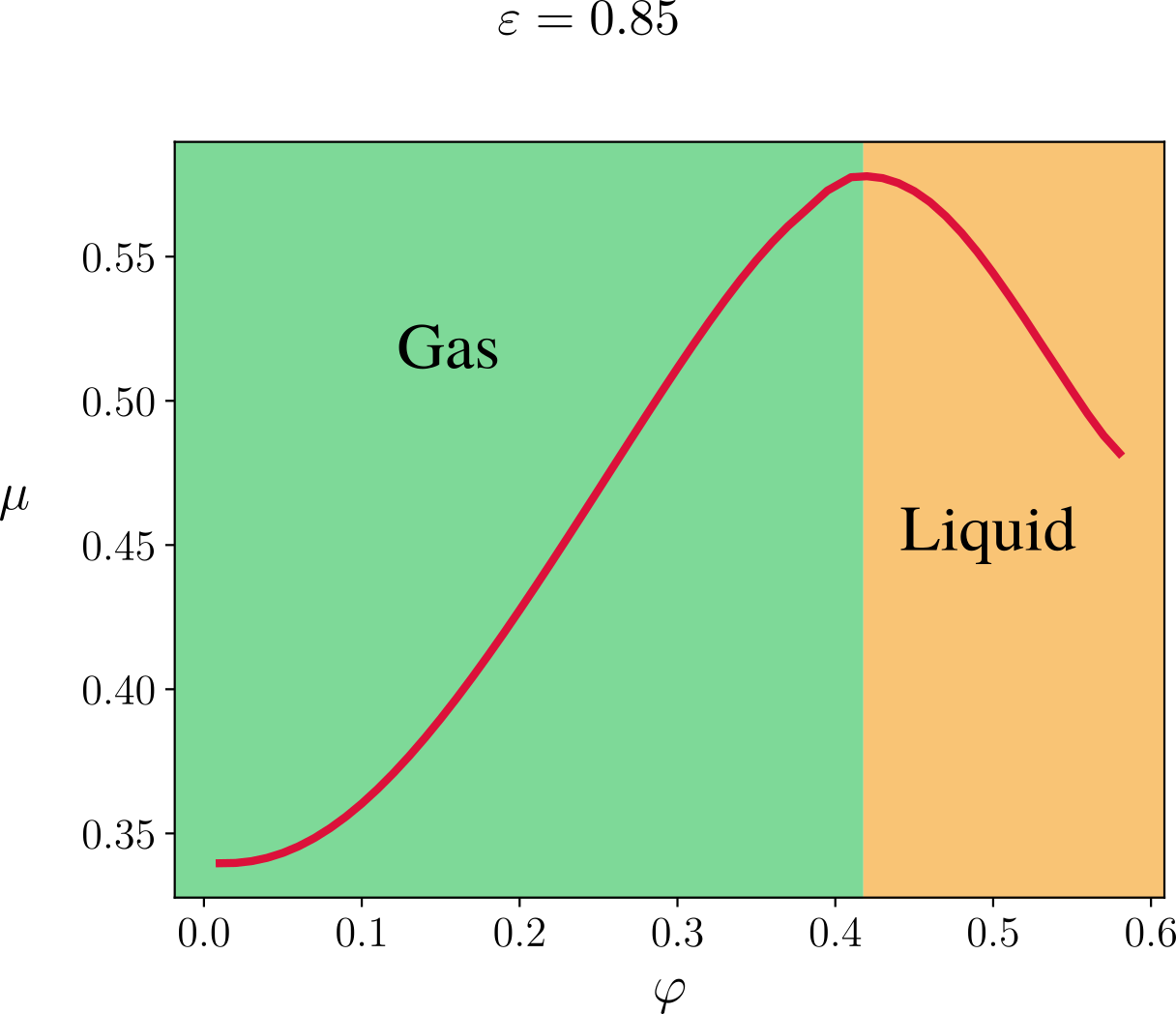}
			\end{center}
			\caption{Evolution of $\mu$ in the GITT model as a function of $\varphi$ for $\varepsilon=0.85$.}
			\label{figMu85}
		\end{figure}

\begin{figure}
			\begin{center}
				\includegraphics[scale=0.55]{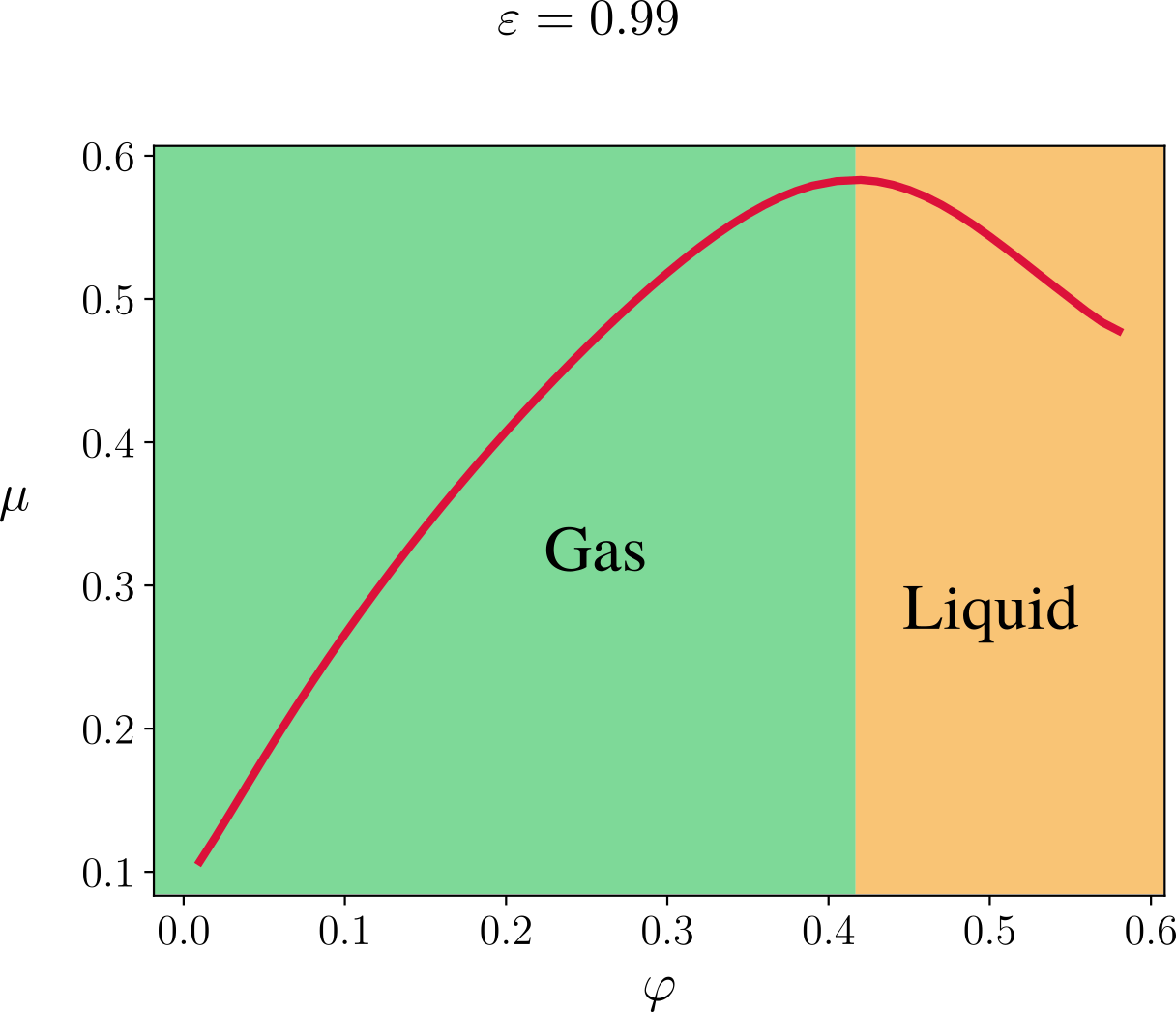}
			\end{center}
			\caption{Evolution of $\mu$ in the GITT model as a function of $\varphi$ for $\varepsilon=0.99$.}
			\label{figMu99}
		\end{figure}

		The evolution of the effective friction coefficient $\mu$ as a function of the packing fraction $\varphi$ is depicted on Fig.~\ref{figMu85}
		for $\varepsilon=0.85$ and Fig.~\ref{figMu99} for $\varepsilon=0.99$.

		First, $\mu$ is a decreasing function of $\varphi$ for $\varphi\gtrsim0.40$.
		This corresponds to the $\mu(\mathcal{I})$ regime investigated in previous studies \cite{Coquand20f,Coquand21};
		It is hence identified with the liquid state, in agreement with the definition of a granular liquid given above.

		Then, around $\varphi\simeq0.40$, $\mu$ shows a monotonicity change.
		This effect is very clearly visible and is thus expected to be clearly visible experimentally.
		Given the range of packing fractions generally taken into account \cite{GDR04,Forterre18,Guazzelli18,Tapia19,Fullard19},
		this would demand only a small deviation away from what is known (if the quantitative predictions of the GITT model are correct).

		By comparison between the predictions of the $\mu(\mathcal{I})$ law, and those of the various models of granular gases exposed above,
		we identify this maximum in the $\mu(\varphi)$ curve as the granular liquid-gas transition.
		It is interesting to remark that this definition is based purely on the behavior of macroscopic quantities, what should not surprise us
		too much insofar as this transition cannot be described in thermodynamic terms because of the intrinsically out of equilibrium nature
		of granular fluids.

		Based on the contents of the GITT equations, we can interpret this transition as follows~: in the granular gas state,
		the macroscopic behavior is dominated by the effect of complex collision processes which, even though they include
		physics beyond the simple Stoßzahl ansatz \cite{Resibois75b}, remain processes involving only a few particles,
		namely, the correlation length $\xi_{gas}$ is of the order of a few particles diameters.

		In the liquid state, however, the physics is dominated by collective modes involving many particles at a time, so that $\xi_{liq}\gg\sigma$.
		These modes build up with the density increase as cage effect becomes more and more important, and the motions to escape the cages
		require to take into account an increasingly large number of particles to be described.

		Another way of describing the same phenomenon is to point out that, as the density increases, the structural relaxation time scale $t_\Gamma$
		distinguishes itself from $t_{ff}$, which is nothing but the manifestation of the cage effect (in the gaseous state, $t_\Gamma\propto t_{ff}$,
		while when a plateau is present, the two time scales --- which delimit the plateau on both sides --- drift apart from one another).
		
\begin{figure}
			\begin{center}
				\includegraphics[scale=0.55]{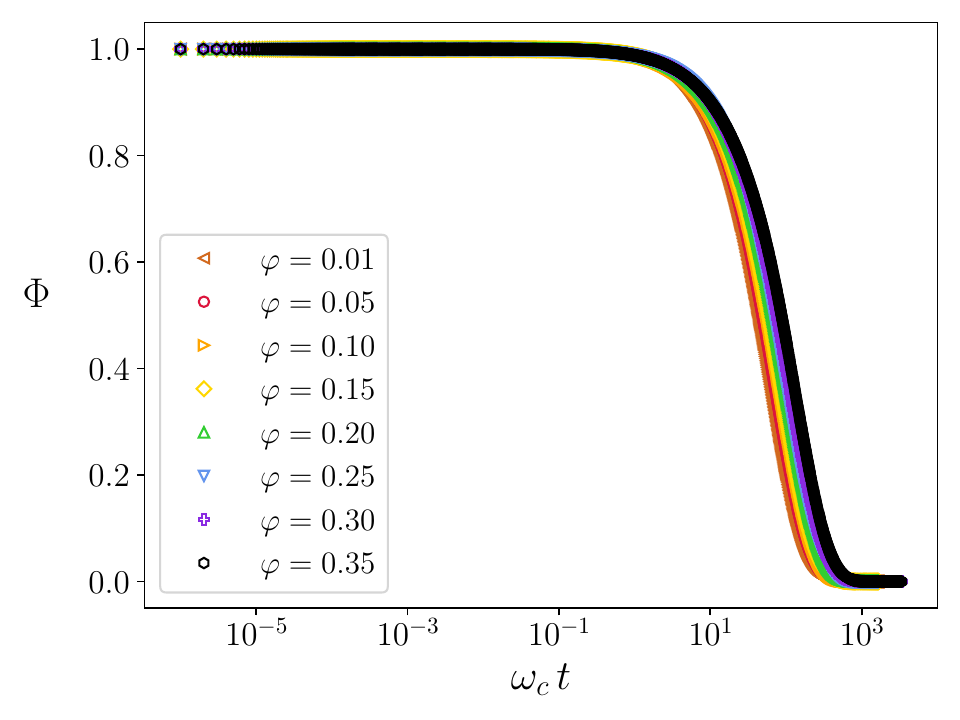}
			\end{center}
			\caption{Evolution of $\Phi$, computed through the GITT equations, in the gas regime.}
			\label{figPhig}
		\end{figure}

		A way to test our time scale interpretation is to have a look at the dynamical structure factor $\Phi_q(t)$ in the gaseous state (we
		already provided a detailed study of its evolution in the liquid regime in \cite{Coquand20g}).
		In the scenario exposed above, $t_\Gamma \propto t_{ff} \propto \omega_c^{-1}$ in the whole range of the gaseous state.
		The evolution of $\Phi$ for packing fractions down to 1$\%$ are represented on Fig.~\ref{figPhig}.
		It can be observed that indeed, this function presents a single decay, at typical scale $\omega_c\,t\simeq1$.
		Our simple toy model picture of the evolution of $\Phi$ is thus consistent with the full solution of the GITT equations.

		But we can test our toy model even further.
		Let us set aside the Bagnold condition for a while.
		What we established so far is that the liquid-gas transition can be described as a crossover from a rheology dependent on three time scales
		to a rheology depending on only two time scales (since two of them basically merge together).
		Indeed, if $t_\Gamma \propto t_{ff}$, then Wi $\propto\mathcal{I}$, Wi$_0/$Wi $\propto \mathcal{I}_0/\mathcal{I}$, and thus~:
		\begin{equation}
			\mu(\text{Wi},\mathcal{I}) = \mu(\mathcal{I}) \simeq \frac{\mu_1}{1+ \mathcal{I}_0/\mathcal{I}} + \frac{\mu_2 - \mu_1}{1+ \mathcal{I}_0/\mathcal{I}}
			 = \frac{\mu_2}{1+ \mathcal{I}_0/\mathcal{I}}\,,
		\end{equation}
		or equivalently, since $\mathcal{I}\propto$ Pe,
		\begin{equation}
		\label{eqMupe}
			\mu(\text{Pe}) = \frac{\mu_2}{1 + \text{Pe}_0/\text{Pe}}\,.
		\end{equation}
		Let us emphasize that this is by no mean obvious since, when allowing physics beyond the Bagnold regime, three time scales (or two dimensionless numbers)
		are required to properly describe the rheology of granular fluids.
		This manifests itself in the formula above by the "disappearance" of the $\mu_1$ term, that was a remnant of the second dimensionless number term.

\begin{figure}
			\begin{center}
				\includegraphics[scale=0.55]{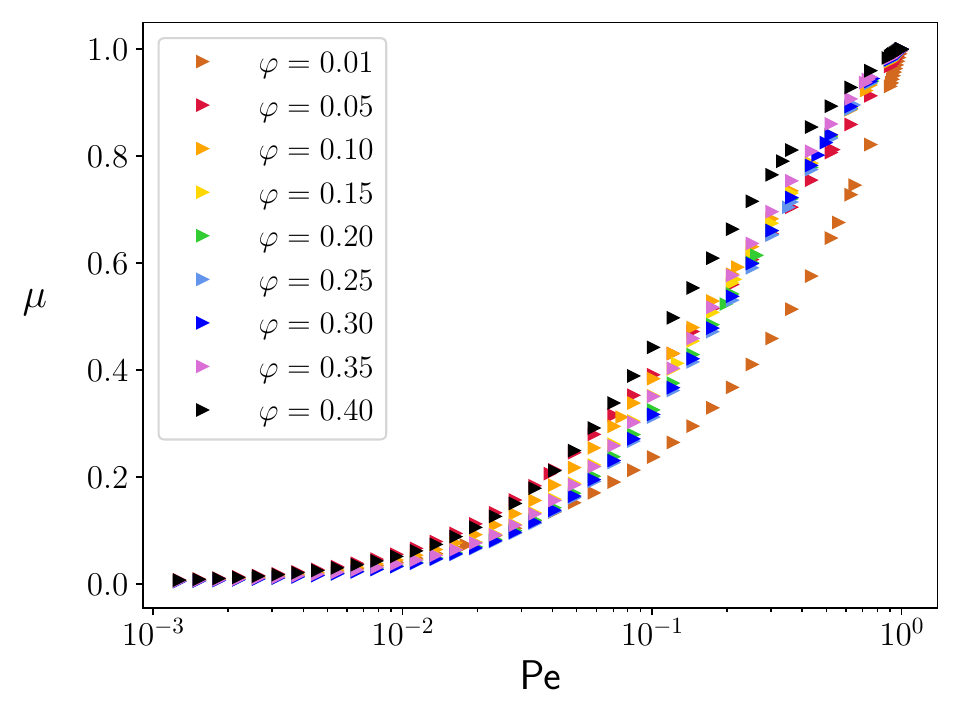}
			\end{center}
			\caption{Evolution of the rescaled $\mu$ in the GITT model as a function of Pe for $\varepsilon=0.99$ and various $\varphi$
			in the gaseous regime.}
			\label{figMas99}
		\end{figure}

\begin{figure}
			\begin{center}
				\includegraphics[scale=0.55]{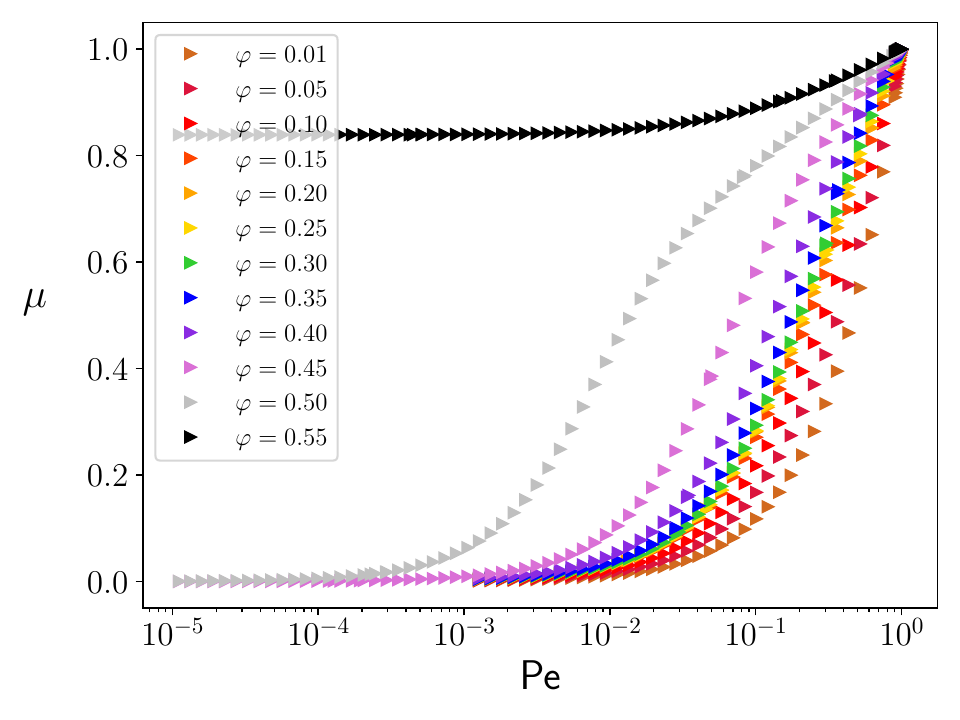}
			\end{center}
			\caption{Evolution of the rescaled $\mu$ in the GITT model as a function of Pe for $\varepsilon=0.85$ and various $\varphi$
			both in the gaseous regime and the liquid regime.}
			\label{figMas85}
		\end{figure}

		The new form of the rheology law Eq.~(\ref{eqMupe}) can also be tested.
		Indeed, if, for each value of $\varphi$, we divide $\mu$ by its value at the maximum Pe (the value Pe$^{\text{max}}$ in the Bagnold regime),
		and Pe by Pe$^{\text{max}}$, all $\mu(\text{Pe})$ curves must collapse onto a single master curve.
		This test has been done on Fig.~\ref{figMas99} for $\varepsilon=0.99$ in the gaseous regime, and on Fig.~\ref{figMas85} in the liquid regime.

		Let us begin with Fig.~\ref{figMas99}.
		It is observed that the collapse is quite convincing, apart for the two extreme values $\varphi=0.01$ and $\varphi=0.40$.
		For $\varphi=0.01$, it is not unexpected that some pathologies may be present as GITT is based on liquid state theory that describes
		matter as a continuous medium, an hypothesis that becomes very dubious at such low packing fractions.
		We will discuss in more details later the dilute limit of the GITT model.
		For $\varphi=0.40$, the granular fluid is very close to the liquid state and the effect of the third time scale begins to show up.

		These effects are even more pronounced on the $\varepsilon=0.85$ curve Fig.~(\ref{figMas85}).
		Here, the breakdown of the master curve profile spreads to higher packing fractions in the dilute limit.
		The evolution of the rescaled $\mu$ into the liquid regime are also presented for comparison.
		For example, on the $\varphi=0.50$ curve, the presence of the third time scale is visible in the form of an additional bump that the
		curve shows at intermediates Pe.
		It must obviously be understood that this feature develops progressively as $\varphi$ is increased, which explains how the
		curves differentiate themselves progressively from the master curve as the granular fluid progresses further into the liquid regime.
		Although the curves seem to show a lot of differences, let us emphasize that the collapse is still good for ${\varphi\in[0.10;0.35]}$.

		The description of the evolution of $\mu$ in the gaseous state as a mere competition between two time scales certainly has its limits,
		as evidenced by the collapse of the curves that is not perfect.
		But let us not forget the context in which this model is analysed.
		First, we are discussing a mere toy model based on a set of (very complex) rheology equations that obviously contain more information
		than the simple formula Eq.~(\ref{eqMupe}).
		However, this equation still holds at the leading order and presents the very interesting property of being simple enough to be
		tested in experiments and numerical simulations.
		Second, as we discussed above, even the best model available to describe the gaseous state do not present a sufficient degree of precision
		to leave a clear quantitative image of the evolution of $\mu$ in the dense gas regime.

		Finally, let us come back to the Bagnold regime.
		On Figs.~\ref{figMas99} and \ref{figMas85}, it corresponds to the end points of the curves (where Pe = Pe$^{\text{max}}$).
		What we can observe is that the function $\mu\big($Pe/Pe$^{\text{max}}\big)$ is increasing.
		On the other hand, as we are going to discuss in details later, Pe$^{\text{max}}$ is a decreasing function of $\varphi$
		(as the gas becomes more and more dilute, a higher shear rate is needed to compensate for the energy lost in collisions since
		advection driven by the boundaries becomes an increasingly inefficient process to influence particle's motion).
		This means that in the gas state, $\mu$ is an increasing function of $\varphi$, what corresponds to the behavior depicted on
		Figs.~\ref{figMu85} and \ref{figMu99}.

	\subsection{GITT in the dilute limit}

		One of the most striking properties of the GITT model in the dilute limit is the absence of monotonicity change that separates the
		dense and dilute gas regimes in the dilute elastic and the sheared Enskog expansion models.
		In order to understand this better, we are going to investigate the GITT model in the dilute limit.

		To do this, we are going to separate two contributions to the shear stress~: $\tau_{xy}^{K}$ comes from the non trivial kernel term,
		and $\tau_{xy}^E$ is the Enskog contribution (see Eq.~(\ref{eqTauGITT})).
		In the dilute limit, the kernel term can be greatly simplified~:
		\begin{widetext}
			\begin{equation}
				\frac{S'\big(q(-t)\big)S'(q)}{S^2(q)}\underset{\varphi\rightarrow0}{=}\frac{576\big[3\,q\cos(q)-3\,\sin(q) + q^2\,\sin(q)\big]^2}{
				q^8}\varphi^2 +O\big(\varphi^3\big)
			\end{equation}
		\end{widetext}
		The dynamical structure factor can be reduced to a simple expression as a function of a diffusion coefficient $D_0$, since, as we have
		established in the previous section, the relaxations of $\Phi$ in the gaseous state are never anomalous.
		\begin{equation}
			\Phi_q(t)\simeq\,\exp\big(-D_0\, q^2\,t\big)\,.
		\end{equation}
		Putting all of this together,
		\begin{equation}
			\begin{split}
				\tau_{xy}^{K} &\underset{\varphi\rightarrow0}{\rightarrow}\frac{\dot{\gamma}T}{60\pi^2\sigma}\int_0^{+\infty}dt\int_0^{+\infty}dq \\
					      &\left[\frac{576\big[3\,q\cos(q)-3\,\sin(q) + q^2\,\sin(q)\big]^2}{q^4}\,\varphi^2\,e^{-2D_0q^2t}\right] \\[0.2cm]
					      &=\frac{288\dot{\gamma}T}{60\pi^2\sigma\,D_0}\,\varphi^2 \\
					      &\times\int_0^{+\infty}dq\frac{\big[3q\cos(q) + (q^2-3)\sin(q)\big]^2}{q^6} \\[0.2cm]
					      &= \frac{12\,\varphi^2}{25\pi}\frac{\dot{\gamma}T}{\sigma\,D_0}\,.
			\end{split}
		\end{equation}
		Then, for dimensional reasons, the diffusion coefficient can be reduced to $D_0=\sigma^2\,\Gamma$, so that~:
		\begin{equation}
			\overline{\tau}_{xy}^K = \frac{\displaystyle \tau_{xy}^K}{nT} 
					       = \frac{12\,\varphi^2}{25\pi}\frac{\dot{\gamma}}{\sigma^3\,\Gamma}\frac{\pi\sigma^3}{6\varphi} 
					       = \frac{2\,\varphi}{25}\frac{\dot{\gamma}}{\Gamma}
		\end{equation}
		Finally, $\dot{\gamma}/\Gamma\propto$ Pe$^{\text{max}}$, which reaches a finite value as $\varphi\rightarrow0$.
		Thus,
		\begin{equation}
			\lim_{\varphi\rightarrow0}\overline{\tau}_{xy}^K=0\,.
		\end{equation}
		On the other hand,
		\begin{equation}
			\lim_{\varphi\rightarrow0}\overline{\tau}_{xy}^E\neq0\,,
		\end{equation}
		otherwise $\mu$ would be 0 in the dilute limit (remember that $\displaystyle\overline{P}\underset{\varphi\rightarrow0}{\rightarrow}1$).
		A similar derivation can be done on the pressure equation.

\begin{figure}
			\begin{center}
				\includegraphics[scale=0.55]{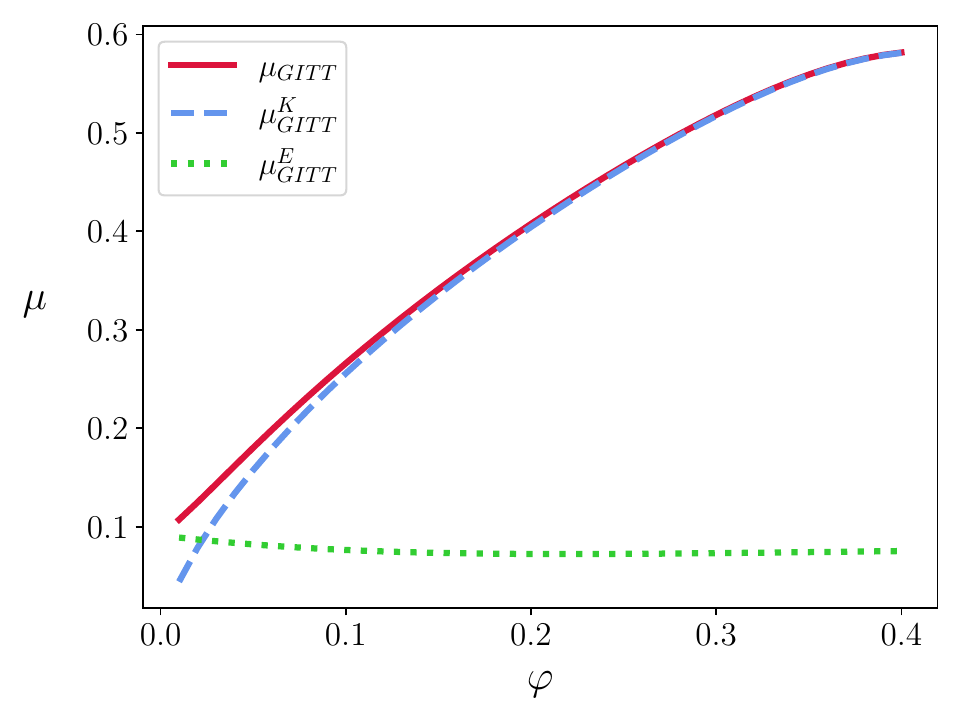}
			\end{center}
			\caption{Evolution of $\mu$ in the GITT model as a function of $\varphi$ for $\varepsilon=0.99$ in the gaseous regime.
			The separate contributions coming from the kernel term $\mu_{GITT}^K$ and the Enskog term $\mu_{GITT}^E$ are
			also represented.}
			\label{figMuSto}
		\end{figure}

		All in all, consistently with the discussion about the splitting of $m_q$ into $m_q^{(1)}$ and $m_q^{(2)}$, the Enskog contribution does dominate
		in the dilute limit.
		To verify this explicitly on the full GITT equations, we have solved independently the problem introducing (i) both contributions to the kernel,
		(ii) only the kernel term, (iii) only the Enskog term.
		The results for $\mu$ are shown on Fig.~\ref{figMuSto}.
		It is observed that, as expected from the asymptotic analysis of the GITT equations, the kernel contribution does die out in the dilute limit,
		whereas the Enskog contribution can be safely neglected out of this regime.

		However, it appears that the domination of the Enskog term occurs only deep into the dilute regime.
		This provides a reasonable explanation for the absence of change of monotonicity of $\mu$ in the dilute limit~: the Enskog term
		is too weak compared to the kernel contribution for $\varphi\simeq0.20$ to lead to any change.
		Its dominating regime only starts in the asymptotic region $\varphi\rightarrow0$, which is not sufficient to enable it to revert the monotonicity
		of the $\mu$ curve (note however that, consistently with what can be expected, the kernel contribution is an increasing function of $\varphi$
		whereas the Enskog contribution is indeed decreasing).

\begin{figure}
			\begin{center}
				\includegraphics[scale=0.55]{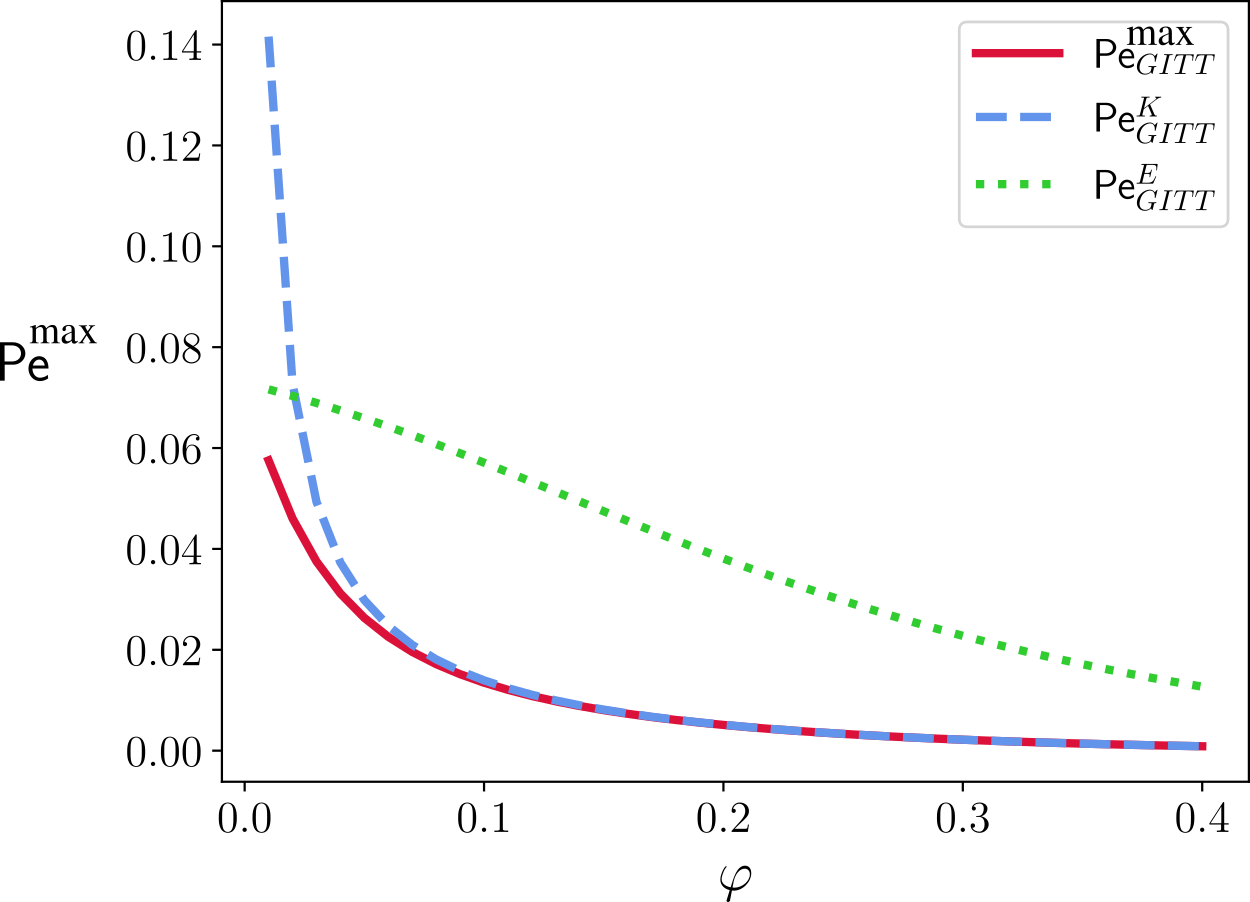}
			\end{center}
			\caption{Evolution of Pe$^{\text{max}}$ as a function of $\varphi$ in the gaseous regime for $\varepsilon=0.99$.
			The evolution for the sole kernel term, and for the sole Enskog term are shown for comparison.}
			\label{figPeSto}
		\end{figure}

		In order to refine our analysis, we now focus on the evolution of the Peclet number Pe$^{\text{max}}$.
		First, according to the power balance equation in the Bagnold regime (\ref{eqPow2}), Pe$^{\text{max}}=\Gamma_d/\overline{\tau}$,
		where we recall that ${\Gamma_d=(1 - \varepsilon^2)/3}$ is a constant.
		The figure \ref{figPeSto} represents the evolution of Pe$^{\text{max}}$ computed for the three separate contributions corresponding to Fig.~\ref{figMuSto}.
		Let us make clear that the procedure consists in solving the full GITT equations for each separate term.
		This means that Eq.~(\ref{eqPow2}) must be solved independently in each case, and Pe$^{\text{max}}$ is not the sum of the kernel and Enskog terms.

		As we discussed above, consistently with physical expectations Pe$^{\text{max}}$ is indeed a decreasing function of $\varphi$ in each of the three cases.
		Then, for the kernel contribution, as $\overline{\tau}\rightarrow0$ in the $\varphi\rightarrow0$ limit, and $\Gamma_d$ is finite, Pe$^{\text{max}}$
		is expected to diverge in the dilute limit, which is what is observed on Fig.~\ref{figPeSto}.

		Finally, Pe$^{E}_{GITT}$ and Pe$^{\text{max}}_{GITT}$ do tend to the same value in the dilute limit, but the difference between the two remains
		significant on the graph, which displays data up to $\varphi=0.01$.
		The global picture emerging from the comparison between the three values of Pe$^{\text{max}}$ confirms the scenario exposed above~:
		For the vast majority of values of $\varphi$ --- note that only the gaseous state is displayed --- down to $\varphi\simeq0.1$, the full
		Pe$^{\text{max}}$ is almost indistinguishable from the kernel contribution alone.
		Only in the very dilute limit do the two curve diverge from each other with signals the onset of the regime in which the Enskog
		contribution becomes important.
		However in practice, the numerical values are such that the regime dominated by the Enskog term, that should agree with the
		sheared Enskog expansion, and therefore trigger a new change of monotonicity in $\mu$, only exists in the asymptotic limit $\varphi\rightarrow0$,
		and the dilute gas regime is never observed.

\begin{figure}
			\begin{center}
				\includegraphics[scale=0.55]{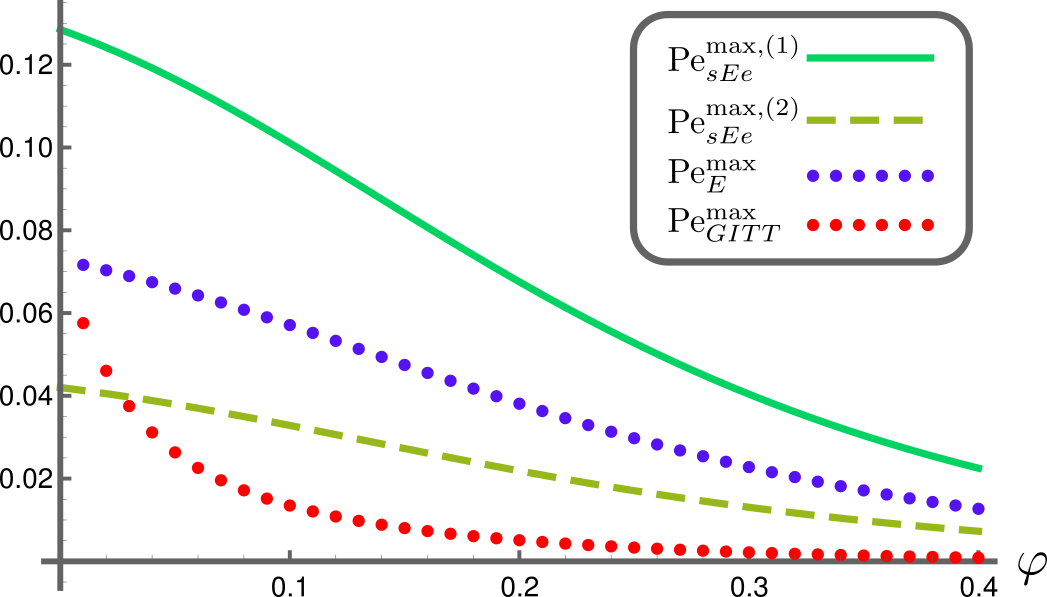}
			\end{center}
			\caption{Evolution of Pe$^{\text{max}}$ as a function of $\varphi$ in the gaseous regime for $\varepsilon=0.99$.
			The GITT values are compared to their equivalent in the sheared Enskog expansion.}
			\label{figPeMath}
		\end{figure}

		A final check of consistency can be conducted comparing the GITT model to the sheared Enskog expansion.
		The results are displayed on Fig.~\ref{figPeMath}.
		On the GITT side, we plotted the full GITT solution as well as the one that only contains the Enskog term.
		For the sheared Enskog expansion, things are a bit more subtle since the power balance equation (\ref{eqPow2}) is never solved self-consistently,
		but truncated in the elastic expansion at next to leading order.
		As a result,
		\begin{equation}
			\frac{4}{5}a^* = \text{Pe}^{\text{max},(1)}_{sEe} \neq \frac{\Gamma_d}{\overline{\tau}_{xy}^{sEe}} = \text{Pe}^{\text{max},(2)}_{sEe}\,,
		\end{equation}
		(the 4/5 prefactor comes from a different definition of the collision frequency).
		With this in mind, the picture depicted on Fig.~\ref{figPeMath} appears clearer~: The value of the Enskog contribution Pe$^{\text{max}}_E$
		falls in between the two estimates from the sheared Enskog expansion, thereby confirming the consistency between the GITT formalism and
		the sheared Enskog expansion.
		When the kernel contribution is added, however, the value of Pe$^{\text{max}}$ obtained by solving the full power balance equation self-consistently
		is not compatible at all with the sheared Enskog expansion model, even in the gaseous state, and for a small inelasticity ($\varepsilon=0.99$).
		These different estimates of the value of the Bagnold shear rate in units of the collision frequency certainly plays a role in explaining
		the numerical differences between the GITT and sheared Enskog expansion models, both at a quantitative and qualitative levels.

\section{Conclusion}

\begin{figure}
		\begin{center}
			\includegraphics[scale=0.5]{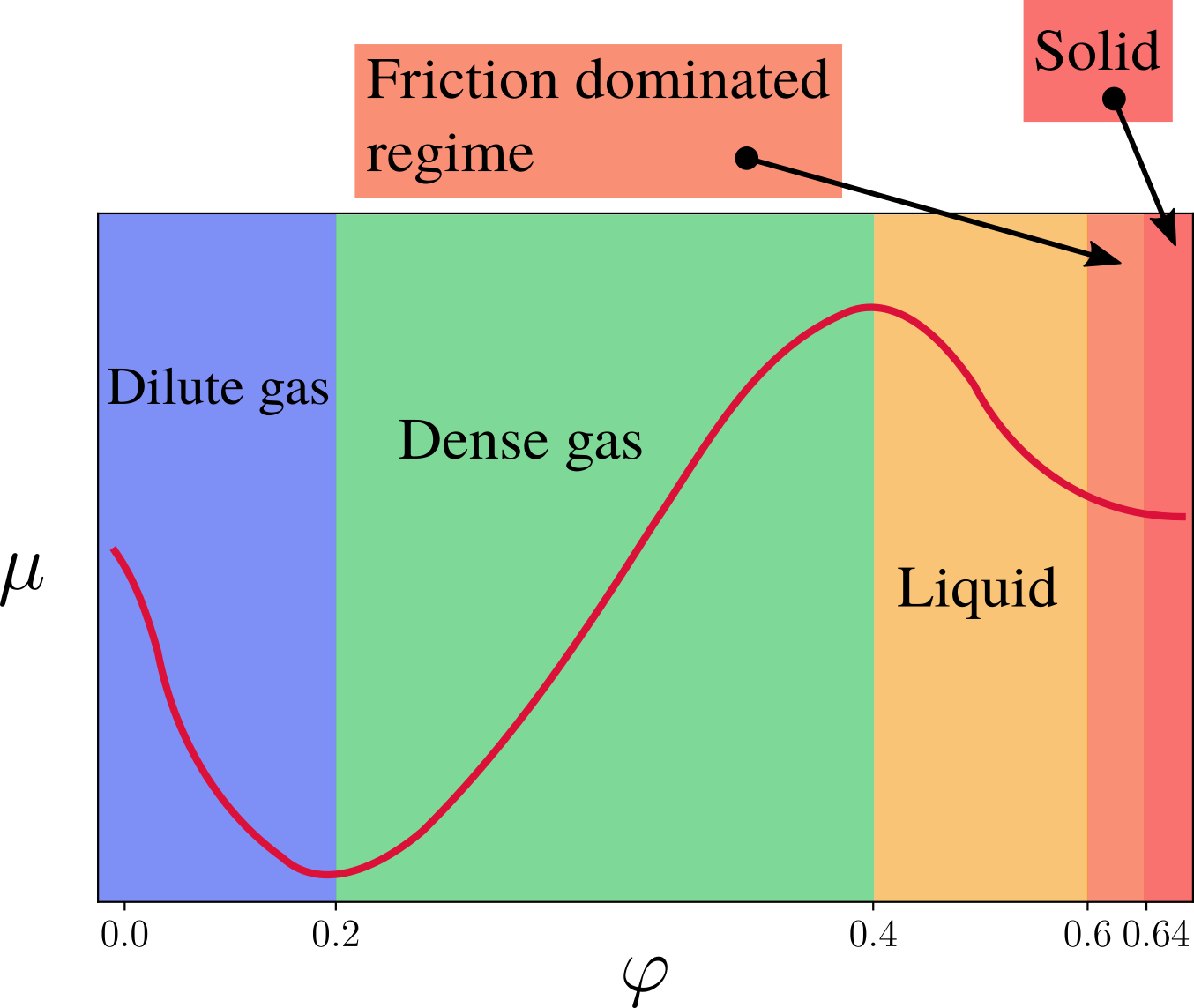}
		\end{center}
		\caption{Sketch of the evolution of $\mu$ as a function of $\varphi$ as predicted by the synthesis of the models studied in this paper.
		The values of $\varphi$ are only orders of magnitude estimated from our numerical result for $\varepsilon=0.85$
		and should not be taken as exact values.}
		\label{figRecap}
	\end{figure}

	To conclude, we have presented a comparison of the best models available to describe the rheology of granular fluids (apart from the friction dominated
	regime).
	The picture that emerges from this analysis qualitatively resembles Fig.~\ref{figRecap}.

	In the dilute limit, the dilute elastic and sheared Enskog expansion models are probably the best adapted models.
	They predict a gaseous state separated in two distinct regimes regarding its rheology~: A first dilute regime where single collision events dominate,
	and a dense gas regime in which more complicated collision processes progressively gain in importance.
	Both regimes are separated by a rupture of the monotonicity of the evolution of $\mu$ with $\varphi$.
	It should be kept in mind however that $\mu$, $\sigma$ and $P$ are macroscopic observables that only make sense at the level of the granular medium
	considered as a whole.
	Experimental tests of this transition should ensure that the granular gas is as homogeneous as possible, and that the notion of global shear stress
	does make sense in such a dilute medium (we think this should be possible if the transition occurs around $\varphi=0.20$ as predicted by the
	Boltzmann inspired models, but can be a problem if the transition is in fact hidden further in the dilute limit).

	When the packing fraction is increased, the granular medium enters the liquid regime.
	The liquid-gas transition is also identified as a rupture of monotonicity of the $\mu(\varphi)$ curve.
	We believe that this prediction should not be too difficult to verify experimentally with our current level of technology since
	the edge of the liquid regime predicted by the GITT model corresponds to the lower limit of the packing fraction range usually tested
	in experiments.
	The liquid regime extends up to the friction dominated regime, which has been well characterized by De Giuli, McElwaine and Wyart \cite{DeGiuli16}.
	In the latter regime, the monotonicity does not change, but the asymptotics of the transition between the liquid and solid states builds up.

	Finally, it appeared throughout our studies that even the best models available today are not sufficient to yield a fully consistent
	picture of the different transitions in the granular fluid at the quantitative level.
	This was to be expected since granular systems are still today notoriously difficult to describe at a theoretical level, due notably to their
	intrinsically dissipative character, and calls for more experiments and numericals simulations to put to the test the prediction presented here.

\section*{Acknowledgements}

	We are particularly grateful to M. Wyart, a discussion with whom started this whole study.
	We are also thankful to A. Santos and V, Garz\'o for their availability and important discussions about their work.
	Finally, we also thank W.T. Kranz and M. Sperl for their help in exploring the depth of the GITT model.

\bibliography{GLG.bib}

\end{document}